\documentclass[pre,onecolumn,notitlepage,longbibliography,amsmath,amssymb,floats,superscriptaddress,nofootinbib,10pt]{revtex4-1}
\usepackage{graphicx,color}
\usepackage{amsmath}
\usepackage{amsfonts, amsbsy}
\usepackage{amssymb}
\usepackage{eqnarray}
\usepackage{bm}
\usepackage{blkarray}
\usepackage{mathtools}
\usepackage{psfrag}
\usepackage{caption}
\usepackage{subcaption}
\usepackage{multirow}
\usepackage{textcomp}
\usepackage{units}
\usepackage{lipsum}
\usepackage[title]{appendix}
\usepackage{soul}
\usepackage{enumitem}
\usepackage{soul}
\usepackage[sort&compress]{natbib}
\usepackage[breaklinks,colorlinks = true,linkcolor = blue,urlcolor  = blue,citecolor = blue,anchorcolor = blue,hyperindex=true, linktocpage=true]{hyperref}
\usepackage{float}

\usepackage[dvipsnames]{xcolor}

\definecolor{nblue}{RGB}{28,130,185}

\newcommand{\be}{\begin{equation}}
\newcommand{\ee}{\end{equation}}

\newcommand{\bg}{\begin{gathered}}
\newcommand{\eg}{\end{gathered}}

\begin{document}

\title{Effective field theory for the superfluid vortex lattice from coset construction}
\begin{abstract}
Guided by symmetry principles, we construct an effective field theory that captures the long-wavelength dynamics of two-dimensional vortex crystals observed in rotating Bose-Einstein condensates trapped in a harmonic potential. By embedding the system into Newton--Cartan spacetime and analyzing its isometries, we identify the appropriate spacetime symmetry group for trapped condensates at finite angular momentum. After introducing a coarse-grained description of the vortex lattice we consider a homogeneous equilibrium configuration and discuss the associated symmetry breaking pattern. We apply the coset construction method to identify covariant structures that enter the effective action and discuss the physical interpretation of the inverse Higgs constraints. We verify that Kohn’s theorem is satisfied within our construction and subsequently focus on the gapless sector of the theory. In this regime, the effective theory accommodates a single gapless excitation--the Tkachenko mode--for which we construct both the leading-order and next-to-leading-order actions, the latter including cubic interaction terms.
\end{abstract}

\author{Aleksander G\l\'{o}dkowski}
\affiliation{Institute of Theoretical Physics, Wroc\l{}aw  University  of  Science  and  Technology,  50-370  Wroc\l{}aw,  Poland}
\affiliation{Nordita, Stockholm University and KTH Royal Institute of Technology, 10691 Stockholm, Sweden}

\author{Sergej Moroz}
\affiliation{Department of Engineering and Physics, Karlstad University, Karlstad, Sweden}
\affiliation{Nordita, Stockholm University and KTH Royal Institute of Technology, 10691 Stockholm, Sweden}

\author{Francisco Pe\~na-Ben\'itez}
\affiliation{Institute of Theoretical Physics, Wroc\l{}aw  University  of  Science  and  Technology,  50-370  Wroc\l{}aw,  Poland}

\author{Piotr Sur\'{o}wka}
\affiliation{Institute of Theoretical Physics, Wroc\l{}aw  University  of  Science  and  Technology,  50-370  Wroc\l{}aw,  Poland}

\maketitle
\flushbottom

\tableofcontents

\section{Introduction}
The experimental realization of Bose–Einstein condensates in dilute atomic gases \cite{doi:10.1126/science.269.5221.198,PhysRevLett.75.3969} has stimulated extensive theoretical and experimental investigations of quantum fluids and solids. Of particular interest is the study of rotating Bose-Einstein condensates, which are realized in analogy to the seminal experiment with a rotating 'bucket' of superfluid helium \cite{PhysRevLett.43.214}. Due to the macroscopic occupation of a single quantum state, the superfluid component cannot undergo rigid-body rotation, but instead carries angular momentum by forming a lattice of quantum vortices \cite{Onsager1949,FEYNMAN195517}. 
Such vortex lattices have been directly observed in several cold atom experiments \cite{PhysRevLett.84.806,PhysRevLett.85.2223,doi:10.1126/science.1060182,PhysRevLett.87.210403,PhysRevLett.89.100403,PhysRevLett.90.170405,PhysRevLett.88.010405,PhysRevLett.91.100402,PhysRevLett.92.040404}.\footnote{More recently, rotating condensates residing in the lowest Landau level (LLL) have been realized at MIT  \cite{doi:10.1126/science.aba7202, mukherjee2022crystallization}, where interaction-induced effects were investigated.} In these setups, angular momentum is imparted onto a trapped gas of cold atoms using a stirring laser that generates a rotating anisotropic potential. After condensation, the system forms a vortex lattice, which equilibrates in the flattened harmonic potential.
When the rotation frequency approaches the trapping frequency, these emergent vortex structures form a triangular lattice that supports transverse elastic modes known as Tkachenko waves \cite{Tkachenko:1965, sonin2014tkachenko}.

Early theoretical work on Tkachenko waves was done in context of rotating superfluid Helium \cite{Sonin1976, volovik1979, Baym1983}.
The emergence of Tkachenko modes in rotating condensates was confirmed experimentally \cite{PhysRevLett.91.110402}, followed by Baym's hydrodynamic theory applicable in the highly-compressible regime \cite{PhysRevLett.91.100402}. Numerous theoretical studies followed thereafter exploring various aspects of the collective excitations in rotating condensates \cite{PhysRevLett.92.060407,PhysRevLett.92.160405,PhysRevLett.92.220401,PhysRevA.71.011603,PhysRevA.72.021606}.

More recently, Tkachenko waves in an infinite two-dimensional vortex crystal were investigated within low-energy effective field theory approach \cite{Watanabe:2013iia,Moroz:2018noc,PhysRevLett.122.235301,10.21468/SciPostPhys.9.5.076, PhysRevB.106.144501, PhysRevResearch.6.L012040, 10.21468/SciPostPhys.17.6.164,PhysRevB.110.035164,PhysRevB.110.024515, manoj2025non}.
In this work, we provide a complementary construction of the effective field theory of vortex crystals by considering the nonlinear realization of the underlying symmetry group \cite{Coleman:1969sm,Callan:1969sn} (see also \cite{PhysRevD.89.045002,PhysRevB.105.205109,10.21468/SciPostPhys.12.5.155,Glodkowski:2024qsf} for applications of the method to many-body systems). Our construction does not rely on a microscopic theory, but rather provides the most general formulation consistent with symmetries. The low-energy degrees of freedom are interpreted as Goldstone modes that transform nonlinearly under the action of spontaneously broken symmetries. We verify that the framework satisfies several phenomenological constraints, such as Kohn's theorem and the expected dispersion of Tkachenko oscillations. Our approach can be generalized to different symmetry groups, including cases of persistent anisotropic stirring \cite{PhysRevLett.86.4443,PhysRevLett.94.150401,PhysRevA.79.011603} and various finite trap geometries \cite{PhysRevLett.92.050403}. 

This manuscript is organized as follows: First, we apply the Newton--Cartan formalism to identify the appropriate spacetime symmetries of the problem. The relevant symmetry group, which we call the magnetic Bargmann symmetry, is constructed and analyzed in detail. Next, we derive the associated effective theory by employing the coset construction method. We discuss the inverse Higgs constraints and verify the validity of the Kohn's theorem. Then, we restrict our attention to the low-energy regime where the effective theory takes a remarkably simple form. Finally, we provide a discussion wherein we put our findings in a broader perspective and compare them with previous results.

\section{Symmetry group for rotating systems in a harmonic trap}\label{sec:symmetryGroup}
In this section, we derive the spacetime symmetries for a rotating system confined in a harmonic potential from first principles. Our analysis is agnostic to whether angular momentum is carried through rigid-body rotation or via the formation of quantum vortices. Therefore, it applies to both normal and Bose–Einstein condensed phases. 

Since Newton--Cartan is the geometric spacetime where Galilean fields propagate \cite{son2025newton, Jensen:2014aia} we apply this framework as a tool to identify the relevant symmetry algebra that we then use to construct the effective field theory for vortex crystals in the subsequent section. 
After a brief review of the Newton--Cartan geometry, we identify the suitable Newton--Cartan background for a system of rotating particles confined in a harmonic potential. We then study the isometries of this spacetime and identify the global symmetries, along with the algebra of the corresponding generators and discuss various limiting cases. 

\subsection{Newton--Cartan spacetimes}
A Newton--Cartan spacetime $(\mathcal{M}, \tau, v, h, A)$ consists of a smooth manifold $\mathcal{M}$ equipped with a symmetric contravariant tensor field $h = h^{\mu \nu} \partial_\mu \otimes \partial_\nu$ that is semi-positive-definite and (singly) degenerate, a nowhere-vanishing clock-form $\tau = \tau_\mu d x^\mu $ in the kernel of $h$, a timelike vector field $v=v^\mu \partial_\mu$ and a $U(1)$ gauge field $A =A_\mu dx^\mu$. In addition, the ''inverse metric'' $h_{\mu\nu}$  is defined, such that 
the following relations hold
\be \label{eq:NCdefinitions}
v^\mu \tau_\mu = 1 \,, \qquad \tau_\mu h^{\mu\nu} = 0 \,, \qquad  h_{\mu \nu} v^\nu = 0\,, \qquad h^\nu_\mu \equiv h_{\mu \rho} h^{\rho \nu} =  \delta^\nu_\mu - v^\nu \tau_\mu\,.
\ee
Notice that these relations are preserved by the field redefinitions
\be 
{v^\prime}^\mu = v^\mu + h^{\mu\nu} \psi_\nu\,, \quad {h^\prime}_{\mu\nu} = h_{\mu\nu}  - \tau_\mu h^\rho_\nu \psi_\rho - \tau_\nu h^\rho_\mu \psi_\rho + \tau_\mu \tau_\nu h^{\rho\sigma} \psi_\rho \psi_\sigma\,,  
\ee
which are usually known as Milne boosts.
With these ingredients we define the non-degenerate tensor $\gamma_{\mu \nu} = \tau_\mu \tau_\nu + h_{\mu \nu}$ and the volume element
\be 
dV = \sqrt{\gamma} d^{d+1} x \,, 
\ee 
where $\gamma = \text{det}(\gamma_{\mu \nu})$, which is a Milne invariant contrary to $\gamma_{\mu\nu}$.
Then, a Galilean field theory can be covariantly coupled to Newton--Cartan geometry as follows \cite{Jensen:2014aia}
\be\label{eq:free}
S[\Psi, \Psi^\dag] = \int dV  \Bigg[ \frac{i}{2} v^\mu \Psi^\dag \overset{\leftrightarrow}{D}_\mu \Psi - \frac{1}{2m}h^{\mu \nu} D_\mu \Psi D_\nu \Psi^\dag  -V( |\Psi|) \Bigg],
\ee
with $D_\mu = \partial_\mu - i A_\mu$. The action \eqref{eq:free} is invariant under diffeomorphism, Milne boost, and $U(1)$ gauge transformations provided the $A_\mu$ transform under Milne boost as 
\be
{A^\prime}_\mu = A_\mu + m h^\nu_\mu \psi_\nu - \frac{1}{2} m \tau_\mu h^{\nu \rho} \psi_\nu \psi_\rho\,.
\ee 
%
Consequently, the field strength $F=dA$ depends on the Milne gauge and hence does not correspond to a physical quantity. 
To identify Milne-independent observables, we note that a thermal many-body system selects a preferred rest frame described by a timelike vector field $\mathcal K = \mathcal K^\mu \partial_\mu$, representing the equilibrium velocity of the medium. Crucially, $\mathcal K$ is a thermodynamic property of the system and does not transform under Milne boosts.
Using this field, one can construct the Milne-invariant combination \cite{Jensen:2014ama}
\be\label{eq:milneInvariantGaugeField}
\tilde A_\mu \equiv A_\mu + m h_{\mu \nu} u^\nu - \frac{m}{2} \tau_\mu h_{\nu \rho} u^\nu u^\rho\,, \quad u^\mu \equiv \frac{\mathcal K^\mu}{\tau_\nu \mathcal K^\nu}\,,
\ee
whose field strength $\tilde F=d \tilde A$ is invariant under local Milne boosts and $U(1)$ gauge transformations thereby encoding the physical electromagnetic curvature. The electric and magnetic field experienced by the system are then given by 
\be \label{eq:electromagneticFields}
\tilde E_\mu = \tilde F_{\mu \nu} u^\nu
 \,, \quad \tilde B = \frac{1}{2} \epsilon^{\mu \nu \rho}  \tau_\mu \tilde F_{\nu \rho}\,.
\ee

Given an infinitesimal transformation $\delta_\chi$ with $\chi =(\xi^\mu \partial_\mu\,, \psi_\mu dx^\mu \,, \Lambda)$ composed of diffeomorphisms, Milne boosts and $U(1)$ gauge transformations parameters, respectively. We define the Lie bracket between two transformation parameters $[\chi_1,\chi_2]$ via 
\be \label{eq:algebraParameters}
[\chi_1, \chi_2]  \equiv \chi_{[12]} = \Big(\mathcal{L}_{\xi_1} \xi_2\,, \mathcal{L}_{\xi_1} \psi_2 - \mathcal{L}_{\xi_2} \psi_1\,, \mathcal{L}_{\xi_1} \Lambda_2 - \mathcal{L}_{\xi_2} \Lambda_1 \Big)\,. 
\ee 
Such that the action of $\delta_\chi$ on the geometric fields $(\tau_\mu, h^{\mu \nu}, v^\mu, A_\mu)$, given by
  \be \begin{split}\label{eq:infitesimal}
  \delta_\chi \tau_\mu &= \mathcal{L}_\xi \tau_\mu = \xi^\nu \partial_\nu \tau_\mu + \tau_\nu \partial_\mu \xi^\nu, \\
\delta_\chi h^{\mu\nu} &= \mathcal{L}_\xi h^{\mu\nu} = \xi^\rho \partial_\rho h^{\mu\nu} - h^{\mu\sigma} \partial_\sigma \xi^\nu - h^{\sigma\nu} \partial_\sigma \xi^\mu, \\
\delta_\chi v^\mu &= \mathcal{L}_\xi v^\mu + h^{\mu\nu} \psi_\nu = \xi^\nu \partial_\nu v^\mu - v^\nu \partial_\nu \xi^\mu + h^{\mu\nu} \psi_\nu, \\
\delta_\chi A_\mu &= \mathcal{L}_\xi A_\mu + m h^\nu_\mu \psi_\nu + \partial_\mu \Lambda = \xi^\nu \partial_\nu A_\mu + A_\nu \partial_\mu \xi^\nu + m h^\nu_\mu \psi_\nu + \partial_\mu \Lambda\,,
\end{split}\ee 
satisfies $[\delta_{\chi_1}, \delta_{\chi_2}] = \delta_{\chi_{[12]}}$.

Therefore, we define an isometry as the set of parameters $\chi_K =(\xi_K^\mu \partial_\mu\,, \psi^K_\mu dx^\mu \,, \Lambda_K)$ that keeps the geometric data invariant \cite{Jensen:2014aia,Matus:2024kyg}. 

\subsection{Galilean fields under rotation}
 In this section, we consider an embedding of a Galilean system at finite density and angular momentum\footnote{We emphasize that how the system accommodates angular momentum is irrelevant for our analysis. We may loosely refer to the system as ''rotating'', although this is somewhat of a misnomer for a Bose–Einstein condensate.} placed in a harmonic trap into the Newton--Cartan spacetime. Then, we study its isometries, compute the associated Lie algebra of conserved charges and discuss various limiting cases. For the remainder of this manuscript, we are working in $2+1$--dimensions.

The isotropic harmonic trap can be naturally embedded into Newton--Cartan spacetime by switching on a finite background gauge field $A_0 = -\frac{m}{2} \omega^2 x^2$, which recovers the trapping potential upon minimal coupling. Accordingly, the associated Newton--Cartan background is
\be\label{eq:harmonicNC}
    \tau_{\mu} = \delta^0_\mu\,, \quad v^\mu = \delta^\mu_0\,, \quad h^{\mu \nu} = \delta^\mu_i \delta^\nu_j \delta^{ij}\,, \quad A_0 = -\frac{m}{2} \omega^2 x^2\,, \quad A_i = 0\,. 
\ee 
which we call Newton-Hooke spacetime. Indeed, we observe that plugging Eq. \eqref{eq:harmonicNC} into Eq. \eqref{eq:free} yields the expected form of the action
\be\label{eq:actionHarmonic}
S[\Psi, \Psi^\dag] = \int dV  \Big[ \frac{i}{2}  \Psi^\dag \overset{\leftrightarrow}{\partial_0} \Psi - \frac{1}{2m}\partial_i \Psi \partial_i \Psi^\dag  -\frac{m}{2} \omega^2 x^2 |\Psi|^2-V(|\Psi|) \Big]\,.
\ee 
This theory is invariant under global $U(1)$ transformations, time translations and spatial rotations, generated by operators
\be\begin{split}
\hat N &= \int d^2x\, |\Psi|^2\,, \\
\hat H &= \int d^2x\, \left[ \frac{1}{2m} |\nabla \Psi|^2 + \frac{m}{2} \omega^2 x^2 |\Psi|^2 + V(|\Psi|) \right]\,, \\
\hat L &= \int d^2x\, \frac{i}{2}   \epsilon_{ij} x_i \Big[\Psi^\dag  \partial_j  \Psi - \Psi \partial_j \Psi^\dag \Big] \,,
\end{split}\ee
whose infinitesimal action is
\begin{equation}
[\hat N, \Psi] = -\Psi\,, \qquad [\hat H, \Psi] = -i \partial_0 \Psi\,, \qquad [\hat L, \Psi] = -i \epsilon_{ij} x_i \partial_j \Psi\,.
\end{equation}

We now switch to the imaginary time formalism and discuss an Euclidean action. Our discussion is mostly based on \cite{Jensen:2014ama} where we refer an interested reader for further details. Since our goal is to study systems at finite density and angular momentum the partition function of the system is 
\be 
\mathcal Z = \text{Tr} \, e^{-\beta (\hat H - \mu \hat N - \Omega \hat L)}\,. 
\ee  
Therefore, evolution in the Euclidean time is generated by the operator $\hat H_\tau = \hat H - \Omega \hat L - \mu\hat N$. This charge generates the symmetry transformation $i[\hat H_\tau, \Psi] =  \mathcal L_{\mathcal K} \Psi + i\mu\Psi $ with the vector field
\begin{equation}
    \mathcal K = \partial_0 - \Omega \epsilon_{ij}x_i \partial_j\,.
\end{equation}
The physical electromagnetic fields experienced by the system (see Eq.~\eqref{eq:milneInvariantGaugeField}) are given by
\be \begin{split} \label{eq:electromagnetic}
\tilde E_i  &= - m (\omega^2 - \Omega^2) x_i \,, \\
\tilde B &=   - 2 m \Omega \,.
\end{split} \ee 
We see that $\Omega \leq \omega$ is required for the system to admit a stable equilibrium state, whereas in the marginal cases $\omega = \Omega$ and $\Omega = 0$ the electric and magnetic field vanish, respectively. 
In what follows, we analyze in detail the symmetry algebra corresponding to each of these cases.
Before doing so, however, we perform a change of Newton–-Cartan frame, consisting of a coordinate transformation followed by a Milne boost.
We emphasize that this is merely a convenient choice of frame and all physical information is already encoded in the Milne-invariant fields Eq.~\eqref{eq:electromagnetic} that determine the symmetry algebra depending on the values of the parameters $\omega$ and $\Omega$.

Note that one has to Wick rotate \eqref{eq:actionHarmonic} along the integral curves of $\mathcal K$ i.e.
\be
\frac{dx_0}{d \lambda} = 1 \,,\qquad \frac{dx_i}{d \lambda} = \Omega \epsilon_{ij} x_j\,.
\ee 
It is then natural to choose the coordinates $(x_0^\prime, x_i^\prime)$ such that $\mathcal K = \frac{\partial }{\partial x_0^\prime}$. This is equivalent to the change of coordinates
\be\label{eq:rotatingFrame}
dx_i = R_{ij}(\Omega t) dx^\prime_j  - \Omega \epsilon_{ik} R_{kj}(\Omega t)  x^\prime_j dx_0 \,, \quad dx_0 = dx_0^\prime\,,
\ee 
where $R_{ij}(\Omega t) = \delta_{ij} \cos{(\Omega t)}-\epsilon_{ij} \sin{(\Omega t)}$ is an $SO(2)$ rotation matrix.
In the new coordinates, both $h_{\mu \nu}$ and $v^\mu$ transform, deviating from their initial simple forms, $h_{\mu \nu} = \delta^i_\mu \delta^j_\nu \delta_{ij}$ and $v^\mu = \delta^\mu_0$. In particular, we now have
\be\begin{split}
h_{\mu \nu} dx^\mu dx^\nu &=dx_i dx_i= dx^\prime_i dx^\prime_i - 2 \Omega \epsilon_{ij} x^\prime_j dx^\prime_i dx_0 + \Omega^2 {x^\prime}^2 dx_0^2 \,, \\
v^\mu \partial_\mu &= \partial_0 = \partial_{0^\prime} - \Omega \epsilon_{ij} x^\prime_j \partial_{i^\prime} \,.
\end{split}
\ee 
Thus, we find that 
\be\label{eq:offdiagonal}
     {v^\prime}^\mu = \delta^\mu_0 - \delta^\mu_i \Omega \epsilon_{ij} x^\prime_j \,, \quad {h^\prime}_{\mu \nu} = \delta_\mu^i \delta_\nu^j \delta_{ij} - \delta_{(\mu}^i \delta_{\nu)}^0 2 \Omega \epsilon_{ij} x^\prime_j + \delta_\mu^0 \delta_\nu^0 \Omega^2 {x^\prime}^2\,. 
\ee 
Let us also note that in these coordinates the projector is no longer purely spatial 
\be 
h^\nu_\mu = \delta^\nu_\mu - v^\nu \tau_\mu = \delta^\nu_\mu - \delta^\nu_0 \delta^0_\mu + \delta^\nu_i \delta^0_\mu \Omega \epsilon_{ij} x^\prime_j\,. 
\ee 
However, since the theory \eqref{eq:free} is covariant under Newton--Cartan transformations, we can bring the metric data back into their initial simple form by performing a suitable Milne boost transformation $\psi_i =  \Omega \epsilon_{ij} x_j$, at the cost of modifying the background gauge field. 

Indeed, after performing a Milne boost transformation $\psi_i =  \Omega \epsilon_{ij} x_j$ we find that $v^\mu$ and $h_{\mu \nu}$ are brought back to their simple forms 
\be
     {v^\prime}^\mu = \delta^\mu_0  \,, \quad {h^\prime}_{\mu \nu} = \delta_\mu^i \delta_\nu^j \delta_{ij} \,. 
\ee 
Meanwhile, the gauge fields transform under Milne boost as
\be
\delta A_0 = \frac{m}{2} \Omega^2 x^2\,, \quad \delta A_i = m\Omega \epsilon_{ij} x_j\,.
\ee 
Therefore, in this frame, the Newton-Cartan geometry data for a rotating system trapped in a harmonic potential is
\be\label{eq:rotatingNC}
    \tau_{\mu} = \delta^0_\mu\,, \quad v^\mu = \delta^\mu_0\,, \quad h^{\mu \nu} = \delta^\mu_i \delta^\nu_j \delta^{ij}\,, \quad A_0 = \frac{m}{2} (\Omega^2-\omega^2) x^2\,, \quad A_i = m\Omega \epsilon_{ij} x_j\,. 
\ee  
Substituting this Newton--Cartan data into \eqref{eq:free} we obtain 
\be\label{eq:action2}
S = \int dV  \Big[ \frac{i}{2}  \Psi^\dag \overset{\leftrightarrow}{\partial_0} \Psi - \frac{1}{2m}|\partial_i \Psi-i m \Omega \epsilon_{ij} x_j \Psi       
       |^2 -\frac{m}{2} \big(\omega^2-\Omega^2\big) x^2 |\Psi|^2-V(|\Psi|) \Big]\,,
\ee 
which agrees with the action considered in \cite{Watanabe:2013iia}. Note that we could have also started directly with \eqref{eq:action2} and embed it into Newton--Cartan framework via \eqref{eq:rotatingNC}. 

Importantly, in \eqref{eq:rotatingNC}, the effects of rotation are incorporated geometrically via a background gauge field. As a result, upon Wick rotation, the Euclidean theory is defined on a thermal circle along the imaginary time direction in the standard way.

From now on, we focus directly on the geometry \eqref{eq:rotatingNC} and study its isometries. The following discussion therefore applies to any Galilean system trapped in a harmonic potential, regardless of whether it is accurately described by the specific field theory \eqref{eq:action2} or not.

\subsubsection{Magnetic Newton-Hooke algebra}
After solving the Killing equations (see appendix \ref{app:rotatingKilling} for computational details) we find that the most general Killing variations $\chi =(\xi^\mu \partial_\mu\,, \psi_\mu dx^\mu \,, \Lambda)$ of the Newton--Cartan background \eqref{eq:rotatingNC} can be parameterized as follows
\be \begin{split}
\chi &= c_0 \Big(- \partial_t\,, 0\,, 0 \Big)  
+\alpha \Big(- \epsilon_{ij} x_i \partial_j\,, 0\,, 0 \Big) + \lambda  \Big(0\,, 0\,, 1 \Big) \,, \\
&+ b^0_j R_{ij}( \Omega t ) \Big( -\frac{ \sin{(\omega t)}}{\omega} \partial_i\,, - \cos{(\omega t)} dx_i - \frac{\Omega}{\omega} \sin{(\omega t)}\epsilon_{ik} dx_k\,,  m\cos{(\omega t)} x_i \Big)  \\
&+ c^0_j R_{ij}( \Omega t )\Big(  -\cos{(\omega t)} \partial_i + \frac{\Omega}{\omega} \sin{(\omega t)}\epsilon_{ik} \partial_k \,, \frac{ (\omega^2 - \Omega^2)}{\omega}  \sin{(\omega t)} dx_i \,, -m\Omega \cos{(\omega t)}\epsilon_{ik}x_k - m\omega \sin{(\omega t) x_i} \Big) \,,
\end{split}
\ee 
where $R_{ij}(\alpha) = \delta_{ij} \cos{(\alpha)}-\epsilon_{ij} \sin{(\alpha)}$ is an $SO(2)$ rotation matrix. These Killing variations are parameterized by seven independent parameters with the associated generators
\be\begin{split}\label{eq:generators}
H &= \Big(- \partial_t\,, 0\,, 0 \Big)\,, \\
N&= \Big(0\,, 0\,, 1 \Big)  \,, \\ 
L &= \Big(- \epsilon_{ij} x_i \partial_j\,, 0\,, 0 \Big)\,, \\
B_i&= R_{ij}( -\Omega t ) \Big( -\frac{ \sin{(\omega t)}}{\omega} \partial_j\,, - \cos{(\omega t)} dx_j - \frac{\Omega}{\omega} \sin{(\omega t)}\epsilon_{jk} dx_k\,,  m\cos{(\omega t)} x_j \Big) \,,\\
P_i &=  R_{ij}( -\Omega t )\Big(  -\cos{(\omega t)} \partial_j + \frac{\Omega}{\omega} \sin{(\omega t)}\epsilon_{jk} \partial_k \,, \frac{ (\omega^2 - \Omega^2)}{\omega}  \sin{(\omega t)} dx_j \,, -m\Omega \cos{(\omega t)}\epsilon_{jk}x_k - m\omega \sin{(\omega t)} x_j \Big)\,.
\end{split}
\ee 
Using \eqref{eq:algebraParameters} it is possible to verify that these generators span an algebra specified with the following brackets 
\be
\boxed{
\begin{array}{rcl@{\quad}rcl}
[L, B_i] & = & \epsilon_{ij} B_j\,, & [L, P_i] & = & \epsilon_{ij} P_j\,, \\[4pt]
[H, B_i] & = & -P_i -2 \Omega  \epsilon_{ij} B_j\,, & [H, P_i] & = & (\omega^2 - \Omega^2) B_i\,, \\[4pt]
[B_i, P_j] & = & \delta_{ij} m N\,, & [P_i, P_j] & = & - 2m\Omega \epsilon_{ij} N\,.
\end{array}
}
\ee
We refer to this structure as the \textit{magnetic Newton–Hooke algebra}, as it generalizes the ordinary Newton–Hooke algebra (see \ref{sec:NewtonHooke})
to include the effects of an effective magnetic field. To the best of our knowledge, this algebra has not been previously written down in this basis, although it was identified by Gibbons and Pope in their study of nonrelativistic electrons placed in a homogeneous external magnetic field and harmonic potential \cite{Gibbons:2010fb}. The particular basis we employ has a conceptual advantage over that used in \cite{Gibbons:2010fb}, as it reduces to the Bargmann algebra in the appropriate limit (see Section~\ref{sec:bargmann}). It therefore provides a natural extension of the notion of space translations and boosts to systems in a finite harmonic potential and background magnetic field.

Having established the general case, corresponding to a system rotating at angular velocity $\Omega$ and with trapping frequency $\omega$, we now proceed to examine several limits resulting in a number of well-known algebras.

\subsubsection{Magnetic Bargmann algebra}\label{sec:magnetic}
First, we consider the limit in which the rotation frequency exactly matches that of the harmonic trap $(\Omega \rightarrow \omega)$. In this case, the centrifugal force exactly cancels the trapping potential\footnote{In this limit, the condensate becomes effectively two-dimensional \cite{PhysRevLett.92.040404,PhysRevLett.90.170405}.} and the resulting Newton--Cartan background geometry,
\be
    \tau_{\mu} = \delta^0_\mu\,, \quad v^\mu = \delta^\mu_0\,, \quad h^{\mu \nu} = \delta^\mu_i \delta^\nu_j \delta^{ij}\,, \quad A_0 =0\,, \quad A_i = m\Omega \epsilon_{ij} x_j\,,
\ee  
describes a system placed in an effective homogeneous magnetic field $B_{\textrm{eff}}=-2m\Omega$. 

The corresponding generators of the isometry transformations can be directly obtained by plugging $\omega=\Omega$ into \eqref{eq:generators}. They take the following form 
\be\begin{split}\label{eq:magneticGenerators}
H &= \Big(- \partial_t\,, 0\,, 0 \Big)\,, \\
N&= \Big(0\,, 0\,, 1 \Big)  \,, \\ 
L &= \Big(- \epsilon_{ij} x_i \partial_j\,, 0\,, 0 \Big)\,, \\
B_i&= \frac{1}{2\Omega} \Big( -\epsilon_{ij} \partial_j\,, 0 \,,    m\Omega x_i \Big) + \frac{1}{2\Omega} R_{ij} (-2 \Omega t)\Big( \epsilon_{jk} \partial_k\,, - 2 \Omega dx_j \,, m\Omega x_j \Big)\,,\\
P_i &=  \Big(   -\partial_i\,, 0\,,-m\Omega \epsilon_{ij} x_j  \Big)\,.
\end{split}
\ee 
These generators obey the following algebra
\begin{equation}\label{eq:magneticBargmann}
\boxed{
\begin{array}{rcl@{\quad}rcl}
[L, B_i] & = & \epsilon_{ij} B_j\,, & [L, P_i] & = & \epsilon_{ij} P_j\,, \\[4pt]
[H, B_i] & = & -P_i -2 \Omega \epsilon_{ij} B_j\,, & [P_i, P_j] & = & -2m\Omega \epsilon_{ij} N\,, \\[4pt]
[B_i, P_j] & = & \delta_{ij} mN\,. & & &
\end{array}
}
\end{equation}
We will refer to \eqref{eq:magneticBargmann} as the \emph{magnetic Bargmann algebra}, as it generalizes the Bargmann group to the case of a finite background magnetic field\footnote{This algebra was also written down in \cite{PhysRevB.111.L161107}.}. This is precisely the centrally extended symmetry algebra of a nonrelativistic particle moving in a homogeneous magnetic field (see Appendix \ref{app:magneticBargmann}). 

In the literature, it is often stated that the presence of a magnetic field breaks boost symmetry, so that the remaining symmetry group consists only of the Hamiltonian, (magnetic) translations, and rotations (see, for example, \cite{PhysRevB.54.5334}). However, we emphasize that this is not the case--there is still a well-defined notion of (magnetic) boost symmetry.


Interestingly, in the fast-rotating limit, the magnetic Bargmann algebra contracts into two invariant subalgebras \eqref{eq:contracted}, one of which is the well-known magnetic translation algebra \cite{PhysRev.134.A1602, PhysRevB.54.5334}. To demonstrate this, we consider the double limit $\Omega \rightarrow \infty$ and $m\to 0$ while keeping the effective magnetic field $B_{\text{eff}}=-2m \Omega$ fixed. Physically, this procedure corresponds to projecting the quantum dynamics onto the lowest Landau level (LLL). 

Before performing the Inönü--Wigner contraction of the algebra we rescale the Hamiltonian $H = \Omega \tilde{H}$ leading to the modified bracket 
\be
[\tilde H, B_i]  = -\tfrac{1}{\Omega} P_i - 2 \epsilon_{ij} B_j\,.
\ee
Applying the contraction, we see that the momentum generator drop out from $[\tilde{H}, B_i]$ in the $\Omega \to \infty$ limit and that the bracket $[B_i, P_j]$ vanishes after sending $m\rightarrow 0$. Therefore, in the fast-rotating LLL limit, the magnetic Bargmann algebra contracts to a direct sum of two invariant subalgebras
\begin{equation}
\begin{aligned}\label{eq:contracted}
[P_i, P_j]      & =  \underbrace{-2 m \Omega}_{ B_{\textrm{eff} }} \epsilon_{ij} N\,,          & \quad [L, P_i]        & = \epsilon_{ij} P_j\,, \\
[\tilde H, B_i] & = -2 \epsilon_{ij} B_j\,,       & \quad [L, B_i]        & = \epsilon_{ij} B_j\,.
\end{aligned}
\end{equation}
In particular, notice that the upper subalgebra is precisely the magnetic translation algebra often employed in the literature to study the LLL dynamics \cite{PhysRevB.54.5334}. 

While the full contracted algebra includes also the magnetic boosts and their nontrivial commutators with the rescaled Hamiltonian, these generators act trivially on the low-energy degrees of freedom in the LLL (see Appendix~\ref{sec:LLL} for an illustrative example). The magnetic translation algebra forms a closed subalgebra containing all generators that act nontrivially within the LLL, and thus correctly captures the relevant symmetry structure at low energies. For this reason, when focusing on LLL dynamics, it suffice to impose only magnetic translation symmetry, as done, for example, in \cite{PhysRevResearch.6.L012040}.

\subsubsection{Newton-Hooke algebra}\label{sec:NewtonHooke}
Next, we consider the non-rotating limit $(\Omega \rightarrow 0)$, which correspond to the Newton--Cartan geometry \eqref{eq:harmonicNC}. This background corresponds to a physical system confined by a harmonic trap. The simplest example of such a system is given by a classical harmonic oscillator $L = \frac{1}{2} m \dot x^2-\frac{1}{2} m\omega^2 x^2$ (see Appendix \ref{app:NewtonHooke}). 

In this case we have the following set of generators 
\be\begin{split}
H &= \Big(- \partial_t\,, 0\,, 0 \Big)\,, \\
N&= \Big(0\,, 0\,, 1 \Big)  \,, \\ 
L &= \Big(- \epsilon_{ij} x_i \partial_j\,, 0\,, 0 \Big)\,, \\
B_i&=  \Big( -\frac{ \sin{(\omega t)}}{\omega} \partial_i\,, - \cos{(\omega t)} dx_i \,,  m\cos{(\omega t)} x_i \Big) \,,\\
P_i &=  \Big(  -\cos{(\omega t)} \partial_i \,, \omega  \sin{(\omega t)} dx_i \,, - m\omega \sin{(\omega t)} x_i \Big)\,.
\end{split}
\ee 
The algebra takes the form of the \textit{Newton-Hooke algebra}, 
\be\boxed{\begin{aligned}\label{eq:newtonHooke}
[L, B_i] &= \epsilon_{ij} B_j\,, &\quad
[L, P_i] & = \epsilon_{ij} P_j\,, \\
[H, B_i] &= -P_i \,, &\quad
[H, P_i] & = \omega^2 B_i \,, \\ 
[B_i, P_j] &= \delta_{ij} m N\,,
\end{aligned}}
\ee 
which is the well-known symmetry algebra of a harmonic oscillator \cite{Gibbons:2003rv,Arratia:2010zz} (see also Appendix \ref{app:NewtonHooke} for some explicit examples of systems exhibiting Newton-Hooke symmetry). 

Interestingly, the Newton–Hooke algebra can be obtained as an Inönü–Wigner contraction of the Anti-de Sitter algebra \cite{Bacry:1968zf,Gibbons:2003rv}. In this case one takes $c \rightarrow \infty$ and $\Lambda \rightarrow 0$ while keeping $\Lambda c^2$ fixed.

\subsubsection{Bargmann algebra}\label{sec:bargmann}
Finally, we consider the flat space limit $(\omega, \Omega\rightarrow0)$. The background geometry becomes that of a flat Newton--Cartan spacetime
\be
    \tau_{\mu} = \delta^0_\mu\,, \quad v^\mu = \delta^\mu_0\,, \quad h^{\mu \nu} = \delta^\mu_i \delta^\nu_j \delta^{ij}\,, \quad A_0 = 0\,, \quad A_i = 0\,. 
\ee  
The isometry generators are 
\be\begin{split}
H &= \Big(- \partial_t\,, 0\,, 0 \Big)\,, \\
N&= \Big(0\,, 0\,, 1 \Big)  \,, \\ 
L &= \Big(- \epsilon_{ij} x_i \partial_j\,, 0\,, 0 \Big)\,, \\
B_i&=  \Big(-t\partial_i\,, -dx_i \,, mx_i\Big) \,,\\
P_i &=  \Big(   -\partial_i\,, 0\,,0  \Big)\,.
\end{split}
\ee 
It is reassuring that the generators are realized on spatial coordinates in the familiar way and that they satisfy the Bargmann algebra 
\be\boxed{\begin{aligned}
[L, B_i] &= \epsilon_{ij} B_j\,, &\quad
[L, P_i] &= \epsilon_{ij} P_j\,, \\
[H, B_i] &= -P_i \,, &\quad [B_i, P_j] &= \delta_{ij} mN\,.
\end{aligned}}
\ee
This is the symmetry group for nonrelativistic massive particles. Similarly to the Newton-Hooke algebra, the Bargmann algebra can be obtained as a low-velocity contraction of the Poincaré algebra \cite{Weinberg:1995mt}.

\section{Effective field theory from spontaneous symmetry breaking}
From the perspective of a modern effective field theorist, the low-energy effective action of a given physical system consists of all operators consistent with the system's symmetries, each accompanied by a phenomenological coefficient. Identifying the complete set of such operators is a nontrivial task, especially when some of the symmetries are \textit{spontaneously broken}, giving rise to Goldstone fields that transform nonlinearly under the action of the broken symmetries\footnote{Note that virtually all condensed matter systems spontaneously break at least some of their symmetries \cite{Nicolis:2015sra}.}. The coset construction method \cite{Callan:1969sn,Coleman:1969sm,Volkov:1973vd,Ivanov:1975zq} (see also \cite{Penco:2020kvy,Naegels:2021ivf,Brauner:2024juy} for a modern introduction to the formalism) offers a systematic procedure for identifying the covariant structures constructed from Goldstone fields, with the symmetry breaking pattern as the only input.

\subsection{The coset formalism}\label{sec:coset}
We now outline the general procedure, starting with the identification of the symmetry generators and the parametrization of the coset space. Consider the symmetry breaking pattern $G \rightarrow H$ where the ground state spontaneously breaks the global symmetry group $G$ down to the subgroup $H$. We denote the generators of the symmetry algebra as follows
\begin{align}
\text{unbroken translations:} \quad & \bar P_\alpha \,, \\
\text{other unbroken generators:} \quad & T_A  \,, \\
\text{broken generators:} \quad & X_a \,.
\end{align}

Since $H$ leaves the ground state invariant, the set of physically inequivalent ground states (the vacuum manifold) is in one-to-one correspondence with the coset space $G/H$. A generic element of the coset space $U \in G/H$ can be parametrized as follows
\be \label{eq:coset}
    U(x, \pi) = e^{x^\alpha \bar P_\alpha} e^{ \pi^a(x) X_a},
\ee
where $x^\alpha$ represent the spacetime coordinates and $\pi^a(x)$ are the Goldstone fields. In the coset parametrization \eqref{eq:coset} we have included the factor $e^{x^\alpha \bar P_\alpha}$, which is needed whenever the broken generators have nontrivial commutation relations with spacetime translations, $[\bar{P}_\alpha, X_a] \neq 0$ \cite{Delacretaz:2014oxa,Brauner:2024juy}.

Under a global transformation $g \in G$, the coset representative transforms as
\begin{equation}\label{eq:transformation}
    g\, U(x, \pi) = U(x^\prime, \pi^\prime)\, h(x, \pi, g)\,, 
\end{equation}
where $h(x, \pi, g) \in H$ is a compensating transformation that is chosen to preserve the coset parametrization \eqref{eq:coset}. Using the symmetry algebra, one can invert the relation \eqref{eq:transformation} to extract the transformed spacetime coordinates $x^\prime$ and Goldstone fields $\pi^\prime$. Importantly, the Goldstone fields transform nonlinearly under the broken symmetries, making it cumbersome to construct invariant combinations directly from their transformation laws.

Instead of working with explicit transformation laws, one identifies covariant building blocks by evaluating the Maurer–Cartan form $\boldsymbol{\omega} \equiv U^{-1} d U$, which takes values in the Lie algebra. It can be decomposed as follows 
\be
\boldsymbol{\omega} = e^\alpha_\mu  \big( \bar P_\alpha + D_\alpha \pi^a X_a + A^B_\alpha T_B \big) dx^\mu\,,
\ee 
where we have introduced coefficients $e^\alpha_\mu$, $D_\alpha \pi^a$ and $A^B_\alpha$, which depend on the Goldstone fields and can be read-off directly from the Maurer-Cartan form. Their transformation properties under $g \in G$ follow from Eq. \eqref{eq:transformation}. In particular, we find  
\cite{Ivanov:1975zq,Penco:2020kvy}
\be 
\begin{split}\label{eq:structures}
    e^\alpha_\mu &\rightarrow
    [\rho(h)]^\alpha_\beta \,  e^\beta_\mu \,, \\
        D_\alpha \pi^a &\rightarrow  \, [\rho(h)]^a_b D_\beta \pi^b [\rho^{-1}(h)]^\beta_\alpha \,, \\
           e^\alpha_\mu  A^B_\alpha T_B &\rightarrow h \left( e^\alpha_\mu  A^B_\alpha T_B \right) h^{-1} + h \partial_\mu h^{-1}\,, \\
\end{split}
\ee 
where $[\rho(h)]^\alpha_\beta$ and $[\rho(h)]^a_b$ are different linear representations of the compensating element $h \equiv h(x, \pi, g) \in H$. 

The coefficients $D_\alpha \pi^a$ are called \textit{covariant derivatives} since they transform covariantly under the action of $G$ in a linear representation of $H$. One can then easily construct $G$--invariant combinations of the Goldstone fields by forming $H$--invariant tensor contractions of these covariant derivatives. 

Similarly, the object $e^\alpha_\mu$ is referred to as the \emph{coset vielbein}. It plays a role analogous to the standard vielbein in gravity, and, in particular, defines the covariant volume element $d^d x \det{e}$ \cite{Ivanov:1975zq,Penco:2020kvy}. Finally, $e^\alpha_\mu  A^B_\alpha T_B$ transforms as an $H$-connection, allowing us to define an $H$--covariant derivative
\begin{equation}
   \nabla^H_\alpha \equiv  E^\mu_\alpha \partial_\mu +  A_\alpha^B T_B\,,
\end{equation}
where $E^\mu_\alpha$ is an inverse to $e^\alpha_\mu$.

Putting the ingredients together, we can write down a $G$--invariant action as
\be 
S[\pi^a] = \int  d^d x \det{e} \, L(D_\alpha \pi^a, \nabla^H_\alpha)\,,
\ee 
where the free indices in $L(D_\alpha \pi^a, \nabla^H_\mu)$ must be contracted in the $H$--invariant fashion. Once the Maurer-Cartan form is evaluated using the symmetry algebra, the building blocks \eqref{eq:structures} can be read-off and the effective action constructed.

\subsection{Inverse Higgs mechanism}\label{sec:inverseHiggsMechanism}
According to Goldstone's theorem, whenever internal symmetries are spontaneously broken in a Poincaré-invariant theory, there exists a massless Nambu–Goldstone mode for each broken generator \cite{PhysRev.127.965}. However, this one-to-one correspondence breaks down when it comes to the breaking of the nonuniform symmetries, which do not commute with the generator of translations. In such cases, the number of gapless Nambu-Goldstone bosons can be smaller than the number of broken generators. This reduction goes under the name of the inverse Higgs mechanism \cite{Ivanov:1975zq}. 

In short, the Inverse Higgs recipe can be summarized as follows. Suppose the symmetry algebra contains a commutation relation of the form
\be 
[\bar P_\alpha, X^\prime] \sim X\,,
\ee 
where $\bar P_\alpha$ is an unbroken translation generator (in the $\alpha$-th direction), and $X$ and $X^\prime$ are broken generators. Furthermore, we assume that $X$ and $X^\prime$ do not transform into one another under the action of the unbroken symmetry group; in other words, they do not belong to the same irreducible representation of the unbroken subgroup $H$. This assumption fails, for example, if $X$ and $X^\prime$ are components of a vector that mixes under unbroken spatial rotations. 

The above conditions are sufficient to ensure that the covariant derivative $D_\alpha \pi$, associated with the broken generator $X$, contains a term linear in the Goldstone field $\pi^\prime$ corresponding to the generator $X^\prime$. One can then consistently impose the constraint $D_\alpha \pi = 0$. This results in an algebraic relation between the Goldstone fields $\pi^\prime$ and $\pi$, allowing for the elimination of $\pi^\prime$ in terms of derivatives of $\pi$. In this way, the Goldstone field $\pi^\prime$ can be systematically removed from the effective theory.

The physical meaning of such an elimination can be understood from two complementary points of view. The first interpretation, due to Low and Manohar \cite{PhysRevLett.88.101602}, is based on the observation that two linearly independent broken generators satisfying $[\bar P_\alpha, X'] \sim X$ do not necessarily generate independent massless fluctuations of the order parameter. In such cases, the elimination can be understood simply as a convenient gauge fixing, which removes the redundant Goldstones (see also \cite{Nicolis:2013sga}).

Depending on the details of the symmetry breaking mechanism\footnote{More precisely, on the representation furnished by the order parameter.}, however, such an interpretation is not always accurate. This brings us to the second interpretation of the inverse Higgs mechanism--namely, that it can be understood as integrating out massive modes to obtain an effective low-energy theory
\cite{PhysRevD.89.085004}. 

In the next section, we will indeed encounter both interpretations in the context of vortex crystals and elucidate their physical origin.

\section{Coset construction for vortex crystals}

In this section, we construct an effective field theory for superfluid vortex crystals using the coset construction formalism. We focus on the regime where the angular velocity of the superfluid matches the frequency of the harmonic trap, $\Omega \to \omega$, such that the spacetime isometry group is the magnetic Bargmann group \eqref{eq:magneticBargmann}. In this limit, the centrifugal force precisely cancels the trapping potential, leading to a homogeneous equilibrium configuration.

In addition to the spacetime symmetries, which are associated with isometries of the background geometry \eqref{eq:rotatingNC}, vortex crystals also enjoy a number of \textit{internal symmetries} that reflect emergent properties of the many-body system. After introducing these internal symmetries, we discuss the symmetry breaking pattern associated with the homogeneous equilibrium state. The relevant low-energy degrees of freedom are the \textit{Goldstone fields}\footnote{Not all of these Goldstone fields correspond to gapless modes, see Sec. \ref{sec:inverseHiggs}.} that transform in a nonlinear fashion under the action of the broken symmetries. Applying the coset construction method, we identify derivative structures that transform covariantly under the broken symmetries and thus constitute the elementary building blocks of the effective action. We then discuss the role of the inverse Higgs constraints and their physical interpretations. We verify that Kohn's theorem is satisfied within our framework. Finally, we focus on the low-energy regime and impose a derivative expansion that allows for a systematic truncation of the effective action. 

\subsection{Coarse-grained picture of vortex crystals}
In Sec. \ref{sec:symmetryGroup}, we established the magnetic Bargmann group \eqref{eq:magneticBargmann} as the spacetime symmetry group for a system of rotating particles confined in a harmonic trap. However, up to this point, we have remained agnostic about the finer details of the system. In particular, we have not yet incorporated the fact that we aim to describe a vortex crystal phase. In this section, we discuss the macroscopic properties of vortex crystals, introduce their coarse-grained degrees of freedom, and examine their symmetries, which we will call \emph{internal symmetries} to emphasize that they originate from the intrinsic properties of the system and to distinguish them from the spacetime symmetries.

Upon coarse graining, the low-energy description of a vortex crystal in $d=2$ spatial dimensions can be formulated in terms of two ($I=1,2$) scalar fields $\phi^I(x,t)$ labelling the vortices. These are the comoving (material) coordinates of the solid \cite{PhysRevLett.94.175301,Dubovsky:2005xd}.

Crystals possess discrete translational symmetries due to spatial periodicity of the underlying lattice $\phi^I \rightarrow \phi^I + a^I$. After coarse-graining, lattice translations become effectively continuous at macroscopic scale and the associated translations are generated by the internal momenta $T_I$, which shift the vortex coordinates 
\be\label{eq:shift}
T_I : \phi^I \rightarrow \phi^I + \xi^I\,,
\ee 
where $\xi^I$ are arbitrary parameters.

While long-scale homogeneity is a generic property of solids, the same cannot be said about isotropy since the underlying crystalline structure introduces anisotropies that persist even on large scales. In the context of vortex crystals, Tkachenko showed that the emergent vortex lattice has an equilateral triangular structure \cite{Tkachenko:1965}, and therefore exhibits a discrete $C_6$ symmetry. This discrete rotational symmetry is realized as a rotation of the vortex coordinates  
\be\label{eq:rotation}
C_6 : \phi^I \rightarrow D^I_J \phi^J
\ee 
where $D^I_J \in SO(2)$ is a matrix that represents a two-dimensional rotation by an angle of $\alpha = n \frac{\pi}{3}$.  

In order to implement the conservation of the $U(1)$ particle number charge $N$ we also introduce a scalar compact field $\theta \equiv \theta(x,t)$. This field represents the superfluid phase and transforms under the $U(1)$ symmetry by  a shift 
\be\label{eq:charge}
N : \theta \rightarrow \theta + \lambda\,.
\ee 
Notice that the bracket $[B_i, P_j]=m\delta_{ij}N$ implies that the $\theta$ field necessarily transforms also under the action of boosts. The explicit transformation property can be readily extracted from \eqref{eq:magneticGenerators}. Therefore, we already see that the full symmetry group does not factorize into a direct product of internal and spacetime symmetry groups. 

So far our setup closely resembles that of a supersolid \cite{PhysRevLett.94.175301}. Vortices, however, exhibit a characteristic noncommutativity in their coordinates \cite{10.1063/1.2425103}. In particular, the internal translations of vortex positions obey the magnetic translation algebra \cite{PhysRevD.89.085004} 
\be \label{eq:magneticTranslation}
[T_I, T_J] =  2 m \Omega \epsilon_{IJ} N \,.
\ee 
This algebra implies that under infinitesimal internal translations \eqref{eq:shift} with parameter $\xi^I$, the phase field $\theta$ transforms as
\be \label{eq:magTransformation}
T_I : \theta \rightarrow \theta + m \Omega \epsilon_{IJ} \xi^I \phi^J\,.
\ee
This concludes our discussion of the coarse-grained degrees of freedom and their internal symmetries.

\subsection{Symmetry breaking pattern}
Here we consider ground state vortex crystal configuration and discuss the resulting symmetry breaking pattern. As we will argue below, the symmetries discussed thus far do not individually preserve the equilibrium configuration, although one can still identify some unbroken combinations of generators. The remaining broken generators give rise to Goldstone fields, which govern the low-energy dynamics around the vortex lattice equilibrium state.

The full symmetry group of superfluid vortex crystals is given by a semidirect product of the spacetime magnetic Bargmann group \eqref{eq:magneticBargmann} and the internal magnetic translation group \eqref{eq:magneticTranslation}. The associated Lie algebra is specified by the following commutation relations 
\be\label{eq:algebra2}\begin{aligned}
[L, B_i] &= \epsilon_{ij} B_j\,, &\quad
[L, P_i] &= \epsilon_{ij} P_j\,, &\quad [H, B_i] &= -P_i -2 \Omega  \epsilon_{ij} B_j\,,\\
[B_i, P_j] &= m\delta_{ij} N\,, &\quad [P_i, P_j] &= - 2m\Omega \epsilon_{ij} N\,, &\quad 
[T_I, T_J] &= 2m\Omega \epsilon_{IJ} N \,.
\end{aligned}
\ee
For a homogeneous vortex crystal equilibrium state at finite chemical potential one can fix
\be \label{eq:equilibrium}
\phi^I_{eq}  = \delta^I_i x^i \,, \quad   \theta_{eq}  = \mu t\,.
\ee 
Equilibrium configuration \eqref{eq:equilibrium} spontaneously breaks all symmetries, both spacetime and internal. This is easily verified by considering symmetry variations of \eqref{eq:equilibrium} under a generic infinitesimal transformation parameterized by $\mathcal{\chi} = \{\tau, c^i, b^i, \alpha, \lambda, \xi^I\}$, which correspond to time and spatial translations, boosts, spacetime rotations, $U(1)$ shifts, and internal translations, respectively\footnote{The explicit form of the symmetry variations is deduced from \eqref{eq:magneticGenerators}.}.

However, there are certain linear combinations of the generators that are unbroken in the sense that they preserve the equilibrium state \eqref{eq:equilibrium}. Indeed, let us consider a symmetry variation $\mathcal{\chi} = \{\tau, c^i, \lambda, \xi^I\}$ associated to infinitesimal time and space translations, $U(1)$ transformations and internal translations. The coarse-grained variables transform as 
\be \begin{split}\label{eq:equilibriumVariations}
\delta_\chi  \phi^I  &= - \tau \partial_t \phi^I  - c^i \partial_i  \phi^I + \xi^I \,, \\
\delta_\chi \theta &= - \tau \partial_t \theta  - c^i \partial_i  \theta - m \Omega \epsilon_{IJ}  \delta^I_i c^i \phi^J + \lambda +  m \Omega \epsilon_{IJ}  \xi^I \phi^J\,.
\end{split}
\ee 
Evaluating the variations on the equilibrium state \eqref{eq:equilibrium} we find that the variations vanish for $c^i = \delta^i_I \xi^I$ and $m\tau = \lambda$. Therefore, we identify the unbroken translation generators $\bar P_i = P_i + \delta^I_i T_I$ and $\bar H = H + \mu N$. There is also a notion of unbroken discrete rotational $C_6$ symmetry, which consists of a simultaneous internal rotation \eqref{eq:rotation} combined with a discrete spacetime rotation that leaves the equilibrium configuration \eqref{eq:equilibrium} invariant. Under the action of the unbroken rotation both internal and spacetime indices transform as vectors. Therefore, from now on we stop distinguishing between the indices $i$ and $I$ (associated with spatial and vortex coordinates, respectively), and denote both by $a$. Furthermore, since the $a$ indices are raised and lowered with the delta metric, we will often switch upper and lower indices freely, with Einstein summation implied.

 In summary, the symmetry breaking pattern is
\be\begin{split}\label{eq:breakingPattern}
\text{Unbroken:}& \quad \bar H = H + \mu N\,, \quad  \bar P_a = P_a + T_a\,, \\
\text{Broken:}& \quad T_a\,,  B _a\,, L\,, N \,.
\end{split}\ee 
In this basis of generators the algebra takes the following form 
\be\begin{aligned}\label{eq:cosetAlgebra}
[T_a, T_b] &= 2m\Omega  \epsilon_{ab} N\,, &\quad  [\bar P_a, T_b] &= 2m\Omega  \epsilon_{ab} N \,, \\
[B_a, \bar P_b] &= m \delta_{ab} N\,, &\quad  [\bar H, B_a] &= - \bar P_a + T_a -2\Omega \epsilon_{ab} B_b \,, \\
[L, \bar P_a] &= \epsilon_{ab} (\bar P_b - T_b)\,, &\quad [L,  B_a] &= \epsilon_{ab} B_b\,.
\end{aligned}
\ee 
Notice in particular that the unbroken spatial translations commute among themselves $[\bar P_a, \bar P_b] = 0$.  
\subsection{Maurer-Cartan form and covariant derivatives}
We now proceed to write down a nonlinear realization of the algebra \eqref{eq:cosetAlgebra} associated with the symmetry breaking pattern \eqref{eq:breakingPattern} using the coset construction formalism. We start by parameterizing the coset space as
\be 
U = e^{t \bar H} e^{x^a \bar P_a} e^{u^a  T_a} e^{\varphi  N} e^{v^a B_a} e^{\gamma L}\,, 
\ee 
where we have introduced the Goldstone fields $\{u^a, v^a, \varphi, \gamma\}$ corresponding to broken internal translations, boosts, $U(1)$ particle number symmetry and spatial rotations, respectively. The Goldstone fields are understood as local functions of the coordinates $x^\alpha = (t,x^a)$ and they realize the broken symmetries in a nonlinear way\footnote{The explicit transformations can be derived from \eqref{eq:transformation}.}. 

In order to compute the Maurer--Cartan form, we will make use of the following identities, which are derived from the algebra \eqref{eq:cosetAlgebra} using the Baker--Campbell--Hausdorff formula,
\be\begin{split}\label{eq:identitiesCoset}
    e^{-v^a B_a} \bar  H 
    e^{v^a B_a} &= \bar H -  v^a \bar P_a + v^a T_a -  2\Omega \epsilon_{ab} v^a B_b + \frac{mv^2}{2}  N \,, \\
       e^{-\gamma L} e^{-v^a B_a} \bar  H 
    e^{v^a B_a}  e^{\gamma L}&= \bar H -  v^a R_{ab}(\gamma)\big(\bar P_b - T_b\big) -  2\Omega \epsilon_{ab} v^a R_{bc}(\gamma)B_c + \frac{mv^2}{2}  N \,, \\
      e^{-u^a T_a} \bar  P_a
    e^{u^a T_a} &= \bar P_a + 2m\Omega \epsilon_{ab} u^b N 
    \,, \\
     e^{-v^a B_a} \bar  P_a
    e^{v^a B_a} &= \bar P_a - m v^a N 
    \,, \\
      e^{-\gamma L} B_a e^{\gamma L} &= R_{ab}(\gamma) B_b\,, \\ 
          e^{-\gamma L} \bar P_a e^{\gamma L} &= R_{ab}(\gamma) \bar P_b + \big(\delta_{ab}-R_{ab}(\gamma)\big)T_b\,, \\
    e^{-u^a T_a} d e^{u^a T_a} &= du^a T_a + m\Omega \epsilon_{ab} du^a u^b N\,, \\ 
      e^{-\gamma L} B_a e^{\gamma L} &= R_{ab}(\gamma) B_b\,.
\end{split}
\ee 
Using identities \eqref{eq:identitiesCoset} it is straightforward to evaluate the Maurer-Cartan form $\boldsymbol{\omega} \equiv U^{-1} d U$. We find
\be 
\boldsymbol{\omega} = \omega_H \bar H + \omega^a_P \bar P_a + \omega^a_T T_a +  \omega^a_B B_a + \omega_N N + \omega_L L\,,
\ee
with the following expressions for the components 
\be \begin{split}\label{eq:components}
\omega_H &= dt \,, \\ 
\omega^a_P &= \big(dx^b - v^b dt \big) R_{ba}(\gamma) \,, \\ 
\omega^a_T &= d\phi^a - \omega^a_P  \,, \\ 
\omega^a_B &= \big( dv^b + 2\Omega  \epsilon_{bc} v^c dt \big) R_{ba}(\gamma)\,, \\ 
\omega_N &= d \varphi +2m\Omega \epsilon_{ab} u^b dx^a +  \frac{mv^2}{2} dt-  m v^a dx^a  + m\Omega \epsilon_{ab} du^a u^b  \,, \\
\omega_L &=d\gamma\,.
\end{split}
\ee 

On the other hand, the Maurer-Cartan form can be decomposed as follows 
\be
\boldsymbol{\omega} = e^\alpha_\mu dx^\mu \big( \bar P_\alpha + D_\alpha u^a T_a + D_\alpha v^a B_a + D_\alpha \varphi N + D_\alpha \gamma L\big)
\ee 
where we have defined $\bar P_\alpha = (\bar H, \bar P_a)$. In the above, $e^\alpha_\mu$ is the coset vielbein whereas the covariant derivatives $D_\alpha u^a$ etc. constitute the invariant building blocks that we can use to construct the effective action (see Sec. \ref{sec:coset}). 

From $\omega_H$ and $\omega^a_P$ we can extract the coset vielbein and its inverse 
\be 
e^0_\mu = \delta^0_\mu \,, \quad e^a_\mu = (\delta^b_\mu - \delta^0_\mu v^b \big) R_{ba}(\gamma) \,, \quad E^\mu_0 = \delta^\mu_0+\delta^\mu_a v^a\,, \quad E^\mu_a = \delta^\mu_b R_{ba}(\gamma)\,.
\ee 
Then, from \eqref{eq:components} we identify the covariant derivatives
\be
\begin{split}\label{eq:covariantDerivatives}
    D_0 u^a &= E^\mu_0 \big( \partial_\mu \phi^a - e^a_\mu \big) = \big(\delta^\mu_0 + \delta^\mu_b v^b \big) \partial_\mu \phi^a = \partial_0 u^a + v^b \partial_b u^a + v^a\,, \\
        D_a u^b &= E^\mu_a \big( \partial_\mu \phi^b - e^b_\mu \big) =  R_{ca}(\gamma)\partial_c  \phi^b-\delta^b_a\,, \\
        D_0 v^a &= E^\mu_0 \big( \partial_\mu v^b + 2\Omega \epsilon_{bc} v^c \delta^0_\mu \big)R_{ba} = \big(\partial_0 v^b + v^c \partial_c v^b+ 2\Omega \epsilon_{bc} v^c\big)R_{ba}(\gamma)  \,, \\
        D_a v^b &= E^\mu_a \big( \partial_\mu v^b + 2\Omega \epsilon_{bc} v^c \delta^0_\mu \big)R_{ba} = \partial_d v^c R_{da}(\gamma) R_{cb}(\gamma)  \,, \\ 
        D_0 \varphi &= E^\mu_0 \big( \partial_\mu \varphi +2m\Omega \epsilon_{ab} u^b \delta^a_\mu +  \frac{mv^2}{2} \delta^0_\mu- m  v^a \delta^a_\mu  + m\Omega \epsilon_{ab} \partial_\mu u^a u^b \big) \,,\\
        &=  \partial_0 \varphi +  v^a\partial_a \varphi - \frac{mv^2}{2}  +  m\Omega \epsilon_{ab} v^a u^b + m\Omega \epsilon_{ab} D_0 u^a u^b \,, \\
         D_a \varphi &= E^\mu_a \big( \partial_\mu \varphi +2m\Omega \epsilon_{ab} u^b \delta^a_\mu +  \frac{mv^2}{2} \delta^0_\mu- m  v^a \delta^a_\mu  + m\Omega \epsilon_{ab} \partial_\mu u^a u^b \big) \\
         &= \big(\partial_b \varphi +2m\Omega \epsilon_{bc} u^c  - m  v^b   + m\Omega \epsilon_{cd} \partial_b u^c u^d \big) R_{ba}(\gamma) \,, \\
         D_0 \gamma &= E^\mu_0 \big( \partial_\mu \gamma \big) = \partial_0 \gamma + v^b \partial_b \gamma \,, \\
         D_a \gamma &= E^\mu_a \big( \partial_\mu \gamma \big) = \partial_b \gamma R_{ba}(\gamma) \,.
\end{split}
\ee 

\subsection{Inverse Higgs constraints}\label{sec:inverseHiggs}
In principle, the symmetry breaking pattern supplied us with 6 candidates for Goldstone fields $\{u^a, v^a, \varphi, \gamma \}$. However, not all of these correspond to independent massless fluctuations (see Sec. \ref{sec:inverseHiggsMechanism}). In fact, for vortex crystals, there is only one gapless Nambu-Goldstone boson. 

In this section, we discuss the implementation of the inverse Higgs constraints and clarify the physical interpretation of the inessential Goldstones. In particular, we argue that the boost and rotational ``Goldstones'' correspond to a gauge redundancy and can therefore be removed via a gauge fixing. On the other hand, the Kohn mode, which is associated with the longitudinal component of the displacement field $u^a$, represents a physical massive excitation \cite{Moroz:2018noc,Watanabe:2013iia}. By imposing the corresponding inverse Higgs constraint, the Kohn mode is removed, and the resulting theory exhibits a single gapless excitation associated with transverse oscillations of the vortex lattice. In this long-wavelength regime, the vortex lattice is effectively incompressible in concordance with the LLL effective theory \cite{PhysRevResearch.6.L012040}.

\subsubsection{Redundant Goldstone fields}

We now proceed to verify that the boost and rotational Goldstones correspond to a gauge redundancy. To this end, let us consider the long-wavelength fluctuations of the order parameters \eqref{eq:equilibrium} generated by the broken generators
\be \begin{split}
\delta \phi^I_{eq}  &= \Big( \gamma(x) L + u^a(x) T_a + v^a(x) B_a + \varphi(x) N \Big) \phi^I_{eq}\,, \\
&= \delta^I_a \Big(  \gamma(x) \epsilon_{ab} x_b+ \frac{1}{2\Omega} \epsilon_{ab} v^b(x) -\frac{1}{2\Omega} \epsilon_{ab} R_{bc}(2\Omega t) v^c(x) + u^a(x) \Big)  \,, \\    
\delta \theta_{eq}  &= \Big( \gamma(x) L + u^a(x) T_a + v^a(x) B_a + \varphi(x) N \Big)   \theta_{eq}\,, \\
&= m\Omega \epsilon_{ab} u^a(x) x_b + \frac{m}{2} x_a v^a(x) + \frac{m}{2} R_{ab}(2\Omega t) x_a v^b(x) + \varphi(x) \,.
\end{split}\ee 
We see that the fluctuations generated by $v^a(x)$ and $\gamma(x)$ are not independent from $u^a(x)$ and $\varphi(x)$ and can be absorbed by the following redefinitions 
\be \begin{split}\label{eq:gaugeFix}
\varphi(x) &= \tilde \varphi(x) - m\Omega \epsilon_{ab} u^a(x) x_b - \frac{m}{2} x_a v^a(x) - \frac{m}{2} R_{ab}(2\Omega t) x_a v^b(x)\,, \\
u^a(x)&= \tilde u^a(x)-\gamma(x) \epsilon_{ab} x_b + \frac{1}{2\Omega} \epsilon_{ab} \Big(  R_{bc}(2\Omega t) v^c(x) -  v^b(x)\Big)\,,
\end{split}
\ee 
such that the most general fluctation is parameterized with $\tilde u_a(x)$ and $\tilde \varphi(x)$ only 
\be 
\delta \phi^I_{eq} = \delta^I_a \tilde u^a \,, \quad \delta \theta_{eq} = \tilde \varphi(x)\,.
\ee 
Therefore, an arbitrary fluctuation with non-vanishing $\gamma(x)$ and $v^a(x)$ can be mapped via the redefinition above to one with $\gamma(x) = v^a(x) = 0$. In other words, by employing the gauge fixing \eqref{eq:gaugeFix}, we can eliminate the redundant Goldstone fields $v^a(x)$ and $\gamma(x)$ in favor of $\tilde u^a(x)$ and $\tilde \varphi(x)$. 

Notice that we have already exhausted our gauge-fixing freedom, as the fluctuations $\tilde u^a(x)$ and $\tilde \varphi(x)$ are manifestly independent. In other words, they correspond to physical modes. However, the longitudinal component of the displacement Goldstone is in fact massive. This is an example of the second interpretation of the inverse Higgs mechanism discussed in Sec. \ref{sec:inverseHiggsMechanism} wherein the inessential Goldstone corresponds to a genuine gapped excitation. In the case of the vortex crystal, the relevant massive mode is the Kohn mode, which has the gap $2\Omega$.

\subsubsection{Implementing the constraints}
Having established that the boost and rotational Goldstones are redundant, we now apply the inverse Higgs mechanism to eliminate them in a way that is consistent with the symmetries \eqref{eq:cosetAlgebra}. We then turn to the inverse Higgs constraint for the Kohn mode and discuss its physical implications.

\textit{1. Boost Goldstone.} The first inverse Higgs constraint follows from the bracket $[\bar H, B_a] = - \bar P_a + T_a -2\Omega \epsilon_{ab} B_b$, which implies that we can set 
\be 
  D_0 u^a =  \partial_0 u^a +  v^a +  v^b \partial_b u^a = 0\,,
\ee 
in order to remove the boost Goldstone from the theory in terms of the displacement field $u^a$. 

To solve for the boost Goldstone it is helpful to express the constraint in terms of the comoving coordinates $\phi^a = x^a + u^a$, as follows
\be \label{eq:velocity}
  D_0 u^a =  \partial_0 \phi^a +  v^b \partial_b \phi^a = 0\,.
\ee 
The above equation is the well-known definition of the velocity of the medium. This is because the comoving coordinates do not change along the flow i.e. if $x(t)$ is the trajectory of the fluid parcel labelled by $\phi(x,t)$ then $\frac{d \phi(x(t),t)}{dt} =0$. Thus, the boost Goldstone can be identified with the velocity of the medium. 

The most general solution to \eqref{eq:velocity} can be found as follows. Since at each time $t$, the map $\phi_t : \mathbb R^2 \rightarrow \mathbb R^2$, which assigns $\phi_t : x^a \rightarrow \phi^a(x^i,t)$ is a diffeomorphism it is always possible to invert it via $x_t: \phi^a \rightarrow x(\phi,t)$. By doing so we simply switch between the Lagrangian and Eulerian descriptions of the fluid flow familiar in the context of fluid dynamics. Therefore, we can solve the constraint directly by inverting it
\be 
v^a = - {(\partial \phi^{-1})}^a_b \partial_0 \phi^b \,. 
\ee 
For $\phi^a = x^a + u^a$ we have $x^a = \phi^a - u^a$ hence ${(\partial \phi^{-1})}^a_b = \delta^a_b - \partial_b u^a + \dots$ and consequently we find 
\be \label{eq:boostIHC}
v^a =- \partial_0 u^b \big( \delta^a_b - \partial_b u^a + \dots \big) =  -\partial_0 u^a + \partial_0 u^b \partial_b u^a + \dots 
\ee 
A more elegant way to resolving the constraint is to define $v^\mu = (1, v^i)$ so that the constraint becomes
\be
v^\mu \partial_\mu \phi^I = 0 \,.
\ee 
The solution is given by the velocity of the lattice points\footnote{This velocity has a clear geometrical interpretation as a current of the lattice points since it is given as a Hodge dual of the volume form defined on the vortex space $v = \star \text{vol}(\mathcal M_\phi) = \star \frac{1}{2} \epsilon_{IJ} d\phi^I \wedge d\phi^J$.}
\be 
v^\mu = \frac{1}{|\partial \phi|} \epsilon^{\mu \nu \rho} \epsilon_{IJ} \partial_\nu \phi^I \partial_\rho \phi^J\,, 
\ee 
where $|\partial \phi| = \epsilon^{ij} \epsilon_{IJ} \partial_i \phi^I \partial_j \phi^J$.

\textit{2. Rotational Goldstone.} From $[L,\bar P_a] = \epsilon_{ab} (\bar P_b - T_b)$ it follows that the rotational Goldstone can be consistently removed from the theory by imposing the inverse Higgs constraint $\epsilon_{ab} D_a u^b =0$. This constraint fixes the rotation matrix in terms of the displacement Goldstone 
\be 
\epsilon_{ab} D_a u^b = \epsilon_{ab} \big(R_{ca}(\gamma) \partial_c \phi^b - \delta^b_a \big) = 0\,, \quad  \epsilon_{ab} \partial_a u_b \cos{\gamma} + (2 + \partial_a u^a ) \sin{\gamma} = 0 \,. 
\ee 
The solution for the rotational Goldstone $\gamma$ can, in turn, be obtained iteratively
\be \label{eq:gammaIHC}
\gamma = -\frac{1}{2} \epsilon_{ab}\partial_a u^b + \frac{1}{4} \epsilon_{ab}\partial_a u^b \partial_c u^c + \dots
\ee 

Interestingly, in certain elastic systems known as micropolar (or Cosserat) solids—wherein the constituents of the medium carry finite intrinsic angular momentum—the rotational Goldstone corresponds to a physical massive excitation \cite{Gromov:2019waa,PhysRevB.105.205109}. In these systems, there exists an additional order parameter associated with the local orientation, denoted $\gamma_{\text{eq}}$, which promotes the rotational Goldstone from a gauge redundancy to a physical mode. It would be interesting to investigate whether such a Cosserat mode could play a role in the dynamics of vortex crystals, perhaps in specific regimes. If so, it ought to be retained in the effective theory. However, the relevance of the Cosserat mode to the vortex crystal problem remains unclear to the authors at present.

\textit{3. Kohn mode.} Finally, the bracket $[\bar P_a, P_b] = -2m\Omega \epsilon_{ab} N$ implies that one can consistently set $D_a \varphi =0$ in order to eliminate the displacement field $u^a$ in favour of the phase fluctuation $\varphi$. Imposing this inverse Higgs constraint removes the Kohn mode, which is a physical excitation with mass $2\Omega$, associated with the center-of-mass motion of the condensate, as guaranteed by Kohn’s theorem \cite{PhysRev.123.1242}. We verify this in later parts of the manuscript, while here we discuss the resolution of the constraint and its physical implications.

The constraint can be solved iteratively as follows
\be \begin{split}\label{eq:IHC}
D_a \varphi &= \big(\partial_b \varphi +2m\Omega \epsilon_{bc} u^c  - m  v^b   + m\Omega \epsilon_{cd} \partial_b u^c u^d \big) R_{ba}(\gamma)  =0 \,, \\ 
u^a &= \frac{1}{2m\Omega} \epsilon_{ab} \partial_b \varphi  + \frac{1}{8m^2\Omega^2} \epsilon_{ab} \epsilon_{cd} \partial_b \partial_c \varphi \partial_d \varphi + \dots 
\end{split}
\ee 
where the ellipsis denotes higher-order derivative corrections which will not be relevant for our purposes. This constraint eliminates the longitudinal fluctuations of the displacement field resulting in an incompressible vortex lattice $\partial_a u^a \simeq 0$\footnote{Here, the $\simeq$  indicates that the condition holds only at long wavelengths, i.e., up to subleading corrections in the derivative expansion established in Sec. \ref{sec:derivativeExpansion}. The exact incompressibility corresponds to the LLL condition \eqref{eq:LLL}.}, discussed in \cite{PhysRevB.106.144501, PhysRevResearch.6.L012040}.
To better understand the physical significance of the inverse Higgs constraint \eqref{eq:IHC}, let us consider the LLL limit ($m \rightarrow 0$, $\Omega\to \infty$ with $m \Omega$ being fixed). In this case, the constraint yields 
\be \label{eq:solutionIHC}
u^a = \frac{1}{2m\Omega} \epsilon_{ab} \partial_b \varphi + \frac{1}{2}\epsilon_{ab} \epsilon_{cd} \partial_b u_c u_d \,,
\ee 
which implies that the map $\phi^I=\delta^I_a (x^a + u^a)$ is area preserving, i.e, 
\be \label{eq:LLL}
\frac{1}{2} \epsilon^{ij} \epsilon_{IJ} \partial_i \phi^I \partial_j \phi^J = 1\,.
\ee
As explained in \cite{PhysRevResearch.6.L012040}, this nonlinear constraint arises in the effective theory approach in the LLL regime. In this case $\phi^1$ and $\phi^2$ are not independent fields, but can be both expressed using a single scalar field $\varphi$.

Going beyond the LLL approximation, we must restore the $mv^a$ term in \eqref{eq:IHC}. In that case, we find that the solution for $u^a$ no longer satisfies the nonlinear area-preserving condition \eqref{eq:solutionIHC} exactly. Nevertheless, it still holds approximately because deviations from \eqref{eq:solutionIHC} are subleading in derivatives (see Sec. \ref{sec:derivativeExpansion} for our derivative counting scheme).

\subsection{Kohn's theorem}
As a first check of our construction we proceed to verify that the Kohn's theorem \cite{PhysRev.123.1242} is satisfied within our EFT. To this aim, we choose not to impose the inverse Higgs constraint $D_a \varphi =0$ in order to retain the Kohn mode and proceed to identify its mass. 

Symmetry-invariant action is given in terms of the $C_6$ rotational invariants constructed from \eqref{eq:covariantDerivatives} and derivatives thereof 
\be 
S[u,\varphi] = \int  d^2 x dt\Big[ \mathcal L \big( D_0 \varphi, D_a \varphi, D_0 v^a,D_a v^b, D_a u^b, E^\mu_0 \partial_\mu, E^\mu_a \partial_\mu \big) \Big]\,.
\ee 
In order to identify the massive modes of the theory we set $k=0$ (since we are not interested in the momentum dependence) and focus on terms, which are quadratic in the fields. Then, the theory takes the following form 
\be 
S[u,\varphi] = \int  d^2 x dt\, \mathcal L \big( D_0 \varphi\,, D_a \varphi, D_0 v^a, \partial_0 \big)\,.
\ee 
Since we are neglecting spatial derivatives, the relevant structures simplify to 
\be
\begin{split}
          D_0 v^a &= \partial_0\big( \partial_0 u^a + 2\Omega \epsilon_{ab}  u^b\big) \,, \\
           D_0 \varphi 
         &= \partial_0 \varphi   -\frac{m}{2} \partial_0 u^a\big( \partial_0 u^a + 2 \Omega \epsilon_{ab} u^b\big)  \,, \\
            D_a \varphi &=   \partial_0 u^a + 2\Omega \epsilon_{ab} u^b  \,. 
\end{split}
\ee 
We notice that $  D_0 v^a = \partial_0   D_a \varphi$ so it is not an independent structure and we can write 
\be 
S[u,\varphi] = \int  d^2 x dt\, \mathcal L \big( D_0 \varphi\,, D_a \varphi\,, \partial_0 \big) \,.
\ee 
The most general quadratic theory contains an infinite number of terms and admits the following series expansion
\be \label{eq:actionExpansion}
S[u,\varphi] = \int  d^2 x dt\Big[ n_0 D_0 \varphi +  \frac{\chi}{2} |\partial_0 \varphi|^2 -\frac{\rho}{2} |  D_a \varphi|^2  + \lambda \epsilon_{ab} D_a \varphi \partial_0 D_b \varphi  
+\dots\Big] 
\ee 
Crucially, since we are interested in the massive sector, we are not imposing any derivative counting scheme on the time derivatives and therefore the higher-order terms are not suppressed by a small parameter and the resulting theory is non-local in time. Nevertheless we will still be able to demonstrate that the theory hosts a massive excitation at $\omega = 2\Omega$ in accordance with the Kohn's theorem. The compliance with the Kohn's theorem is enforced by the magnetic Bargmann symmetry \eqref{eq:magneticBargmann}. 

To this aim, we transform into the Fourier space using the following conventions
\be 
\varphi (x,t) =\int \frac{ d\omega}{\sqrt{2\pi}} e^{-i\omega t} \tilde \varphi (x,\omega)\,, \quad u^a (x,t) =\int \frac{ d\omega}{\sqrt{2\pi}} e^{-i\omega t} \tilde u^a (x,\omega)\,. 
\ee 
Then, we arrive at the following general expression for the action in the Fourier space
\be \label{eq:fourierAction}
S[u,\varphi] = \int  d^2 x d\omega \, \begin{pmatrix}
   \tilde u^a(x,\omega) & \tilde \varphi(x,\omega) 
\end{pmatrix}\begin{pmatrix}
    M_{ab}(\omega) & 0\\
    0 &  M(\omega)
\end{pmatrix}\begin{pmatrix}
   \tilde   u^b(x,-\omega) \\
    \tilde \varphi(x,-\omega) 
\end{pmatrix}\,,
\ee
with $M(\omega) = \omega^2 K_0(\omega)$ and
\be \begin{split}\label{eq:matrixDef}
M_{ab} &=\Big[ K_1(\omega) \omega^2 +K_2(\omega) \big(\omega^2+4\Omega^2\big) - K_3(\omega) 4\omega^2 \Omega \Big] \delta_{ab}\,, \\
&-\Big[ K_1(\omega) 2i\omega\Omega +K_2(\omega) 4i\omega\Omega  -K_3(\omega) i\omega (\omega^2 +4\Omega^2)\Big] \epsilon_{ab}
\end{split}
\ee 
where each $K_i(\omega)$ admits infinite series expansion in the powers of frequency, $K_i(\omega) = \sum_n a_{i,2n} \omega^{2n}$. 

In order to identify the excitations of the theory we demand that the determinant of the block diagonal matrix in \eqref{eq:fourierAction} vanishes,
\be
\text{det}(M_{ab})  M(\omega) = 0\,. 
\ee 
Independently of the $K_i$'s we identify two solutions. First, we find a massless mode that follows from imposing $M(\omega)=0 \rightarrow \omega =0$. This is the only massless excitations in the theory and it is known as the Tkachenko mode. We investigate its dynamics in great detail in the following section. The second (massive) excitation is identified after imposing $\text{det}(M_{ab}) = 0$. This condition is equivalent to 
\be \begin{split}
\text{det}(M_{ab})&=\Big[(K_0(\omega))^2 \omega^2 + (K_2(\omega))^2\big(\omega^2 - 4\Omega^2\big) - (K_3(\omega))^2\omega^2\big(\omega^2 - 4\Omega^2\big) +2K_0 K_2 \omega^2  - 4K_0 K_3 \omega^2 \Omega  \Big] \big(\omega^2 - 4\Omega^2\big) \,.
\end{split}
\ee 
Therefore, we find that independently of $K_i$'s there is a massive excitation satisfying $\big(\omega^2 - 4\Omega^2\big)=0 \rightarrow \omega = \pm2\Omega$. 

Notice that at $\omega = \pm 2\Omega$, the matrix $M_{ab}$ simplifies to
\be 
M_{ab}(\pm 2\Omega)= K(\Omega) \left( \delta_{ab} \mp i \epsilon_{ab} \right)\,, \quad K(\Omega) = 4 \Omega^2 \left[ K_1(2\Omega) + 2 K_2(2\Omega) \mp 4\Omega K_3(2\Omega) \right]\,.
\ee
Since $\text{det}(M_{ab}) = 0$ at this frequency, the matrix is not invertible and admits a nontrivial null vector. To determine the polarization of the associated mode, we solve the condition
\be
M_{ab}(\pm 2\Omega)\, u_b = 0\,,
\ee
which yields
\be
\frac{u_x}{u_y} = \pm i\,,
\ee
indicating that the mode is circularly polarized. This is the celebrated Kohn mode whose excitation frequency at zero momentum is exactly twice the rotation frequency.

We emphasize that the presence of the massless Tkachenko mode and the massive Kohn mode is guaranteed by the magnetic Bargmann symmetry \eqref{eq:magneticBargmann}, which in turn determines the form of the covariant derivatives \eqref{eq:covariantDerivatives}. It is therefore independent of the microscopic details of the system. 

\subsection{Effective field theory of the Tkachenko mode}\label{sec:gapless}
In this section, we construct an effective field theory action, which governs the long-wavelength dynamics in superfluid vortex crystals. We focus on the low-energy regime and integrate out all the massive (UV) modes including the Kohn mode. In other words, we impose all three possible inverse Higgs constraints and write down a minimal nonlinear relization of the symmetry breaking pattern \eqref{eq:breakingPattern} in terms of the Tkachenko mode--a single massless Nambu-Goldstone boson with quadratic dispersion relation $\omega \sim k^2$. Moreover, we construct the next-to-leading order action that contains nonlinear cubic terms in the Goldstone field and identify a new interaction term that is allowed by the symmetries.

\subsubsection{Derivative expansion}\label{sec:derivativeExpansion}
In the low-energy regime, the mass of the Kohn mode $ 2\Omega$ provides a high-energy scale, which plays the role of a UV cutoff. This enables a well-defined derivative expansion in temporal derivatives $\mathcal{O}(\partial_t) \sim \frac{1}{\Omega}$. On the other hand, the length of the system $L$ sets the scale necessary to define the derivative expansion in spatial derivatives $\mathcal{O}(\nabla) \sim \frac{1}{L}$. 

Anticipating the low-energy dispersion relation $\omega \sim k^2$, we postulate the counting $\mathcal O(\partial_t) \sim \mathcal{O}(\nabla^2) \sim \mathcal{O}(\epsilon^2)$, where $\epsilon \ll 1$ is a small parameter. Accordingly, we assign the following orders to the Goldstone fields\footnote{The appropriate order hierarchy for the Goldstone fields can be inferred from the inverse Higgs constraints \ref{sec:inverseHiggs}.}  
\be
\varphi \sim \mathcal O(\epsilon^{-1})\,, \quad u^a \sim \mathcal O(\epsilon^0)\,, \quad \gamma \sim \mathcal O(\epsilon)\,, \quad v^a \sim \mathcal O(\epsilon^2)\,.
\ee

Applying this counting to the covariant derivatives Eq. \eqref{eq:covariantDerivatives}, and neglecting structures which contribute to the effective action at order $\mathcal{O}(\epsilon^4)$ or higher, we are left with two symmetry-invariant combinations 
\be  \begin{split}\label{eq:covariantLowEnergy}
 D_a u^b &= \partial_a u^b + \gamma \epsilon_{ab}  + \gamma  \epsilon_{ac} \partial_c u^b - \frac{\gamma^2}{2} \delta_{ab} + \mathcal{O}(\epsilon^3) \,, \\
 D_0 \varphi &= \partial_0 \varphi + v^a \left( \partial_a \varphi + m\Omega \epsilon_{ab} u^b \right) + \mathcal{O}(\epsilon^4)\,.
 \end{split}
\ee 

Using the inverse Higgs constraints Eqs. \eqref{eq:boostIHC}, \eqref{eq:gammaIHC}, and \eqref{eq:IHC}, all Goldstone fields can be expressed in terms of a single scalar Goldstone $\varphi$, and organized into a perturbative expansion
\be \begin{split}
u^a &= \underbrace{ \Theta  \tilde \partial_a \varphi}_{\mathcal{O}(\epsilon^0)}  \, + \, \underbrace{\frac{\Theta^2}{2}   \tilde \partial_a \partial_b \varphi \tilde \partial_b \varphi }_{\mathcal{O}(\epsilon)}  \, + \,\, \mathcal{O}(\epsilon^2)\,, \\
v^a &= - \underbrace{\Theta  \partial_0 \tilde \partial_a \varphi}_{\mathcal{O}(\epsilon^2)} \, + \, \underbrace{ \frac{\Theta^2}{2} \left( \partial_0 \tilde \partial_b \varphi \partial_b \tilde \partial_a \varphi - \tilde \partial_b \varphi  \partial_0 \partial_b \tilde \partial_a \varphi \right)}_{\mathcal{O}(\epsilon^3)}  \,  + \, \mathcal{O}(\epsilon^4)\,, \\
 \gamma &= \underbrace{\frac{\Theta}{2} \partial^2 \varphi}_{\mathcal{O}(\epsilon)} + \underbrace{\frac{\Theta^2}{4} \partial^2 \partial_a \varphi \tilde \partial_a \varphi}_{\mathcal{O}(\epsilon^2)} + \mathcal{O}(\epsilon^3)\,, 
\end{split}
\ee 
where $\Theta = \frac{1}{2m\Omega}$ and $\tilde \partial_a = \epsilon_{ab} \partial_b$. Finally, plugging these expressions into the covariant derivatives \eqref{eq:covariantLowEnergy} we identify the following low-energy building blocks   
 \be
\begin{split}\label{eq:defects}
       D_a u^b &= \underbrace{\Theta \left(  \partial_a \tilde \partial_b \varphi  + \frac{\epsilon_{ab}}{2} \partial^2 \varphi  \right) }_{\mathcal O(\epsilon)} \, + \,   \underbrace{\frac{\Theta^2}{2} \left[ \partial_a \left( \tilde \partial_b \partial_c \varphi \tilde \partial_c \varphi \right) + \frac{\epsilon_{ab}}{2} \partial^2 \partial_c \varphi \tilde \partial_c \varphi  + \partial^2 \varphi \left(\tilde \partial_a \tilde \partial_b \varphi - \frac{\delta_{ab}}{4} \partial^2 \varphi \right) \right] }_{\mathcal O(\epsilon^2)} \, + \, \mathcal{O}(\epsilon^3) \,, \\
         D_0 \varphi 
         &= \underbrace{\partial_0 \varphi}_{\mathcal O(\epsilon)}  \, - \, \underbrace{\frac{\Theta}{2} \partial_0 \tilde \partial_a  \varphi \partial_a \varphi}_{\mathcal{O}(\epsilon^2)} \, + \, \underbrace{\frac{\Theta^2}{4} \partial_0 \partial_a \partial_b \varphi \tilde \partial_a   \varphi \tilde \partial_b \varphi }_{\mathcal{O}(\epsilon^3)}\, + \, \mathcal{O}(\epsilon^4)\,.
\end{split}
\ee 
We emphasize that we have explicitly listed only those covariant derivatives that are relevant to the subsequent analysis, omitting higher-order structures. The second term in $D_0 \varphi$ is a total derivative that accounts for the Berry phase acquired by the vortices. For regular field configurations, the expression for $D_0 \varphi$ can, up to total derivatives, be rewritten as
\be
D_0 \varphi \simeq  \underbrace{ \frac{\Theta^2}{4} \partial_0 \varphi \left[ (\partial_a \partial_b \varphi)^2 - (\partial^2 \varphi)^2 \right]}_{\mathcal{O}(\epsilon^3)}  + \, \mathcal{O}(\epsilon^4)\,,
\ee
where the symbol $\simeq$ denotes equivalence up to total derivative terms. Similarly, we find
\be 
(D_0 \varphi)^2 \simeq \underbrace{(\partial_0 \varphi)^2}_{\mathcal{O}(\epsilon^2)}  + \, \mathcal{O}(\epsilon^4) \,,  \quad (D_0 \varphi)^3 \simeq \underbrace{(\partial_0 \varphi)^3}_{\mathcal{O}(\epsilon^3)}  + \, \mathcal{O}(\epsilon^5) \,.
\ee 

Having identified the appropriate low-energy structures, it is now straightforward to construct the effective action up to leading $\sim \mathcal{O}(\epsilon^2)$ and next-to-leading $\sim \mathcal{O}(\epsilon^3)$ order in the perturbative expansion. 
\subsubsection{Leading-order action}
The leading order corresponds to truncating the action at second order in the expansion, resulting in a quadratic theory
\be \label{eq:actionTkachenko}
S_2[\varphi] = \int d^3 x \Big[ n_0  D_0 \varphi + \frac{\chi}{2} (D_0 \varphi)^2   - E^{(2)}(\partial u) + \mathcal{O}(\epsilon^3)\Big]\,,
\ee 
where $E^{(2)}(\partial u) = \lambda^{abcd} D_a u^b D_c u^d$ is the elastic strain energy with $\lambda^{abcd}= \lambda_1 \delta_{ab} \delta_{cd} + \lambda_2 \delta_{a\langle c} \delta_{d \rangle b}$. Using \eqref{eq:defects}, we find that it admits the following expansion in terms of the phase field 
\be\begin{split}
E^{(2)}(\partial u) &= \frac{\Theta^2 \lambda_2}{2}\big( 2(\partial_a \partial_b \varphi)^2-(\partial^2 \varphi)^2 \big) + \frac{\Theta^3 \lambda_2 }{4} \Big[ \left( \partial_a \tilde \partial_b \varphi + \tilde \partial_a \partial_b \varphi \right) \partial_a \left( \tilde \partial_b \partial_c \varphi \tilde \partial_c \varphi \right)  \Big]  \, + \, \mathcal{O}(\epsilon^4)\,, \\ 
&\simeq \frac{\Theta^2 \lambda_2}{2}(\partial^2 \varphi)^2  + \mathcal{O}(\epsilon^4)\,.
\end{split}
\ee

Therefore, up to a total derivative, \eqref{eq:actionTkachenko} is equivalent to the Lifshitz theory 
\be \label{eq:lifshitz}
S_2[\varphi] \simeq \int d^3 x \Big[ \frac{\chi}{2} (\partial_0 \varphi)^2   - \frac{\Theta^2 \lambda_2}{2} (\partial^2 \varphi)^2 + \mathcal{O}(\epsilon^3)\Big]\,.
\ee 
This theory accommodates a single massless excitation--the Tkachenko mode--with a quadratic dispersion relation, \( \omega = \frac{\Theta^2 \lambda_2}{\chi} k^2 \). Accordingly, the minimal realization of the symmetry breaking pattern \eqref{eq:breakingPattern} is a real scalar field theory exhibiting Lifshitz scaling symmetry.

Note that we have assumed regular field configurations, thereby neglecting topological defects. In the presence of such defects, one must retain the Berry term and use the covariant structures defined in \eqref{eq:defects}, see \cite{10.21468/SciPostPhys.17.6.164,PhysRevB.110.035164} for related discussions.

\subsubsection{Next-to-leading order action}
Expanding the action up to the third order in the derivative expansion yields a cubic action where interaction terms appear. It takes the following form
\be 
S[\varphi] = \int d^3 x \Big[ n_0  D_0 \varphi + \frac{\chi}{2} (D_0 \varphi)^2   - E^{(2)}(\partial u) + g_1 (D_0 \varphi)^3 + g_2 D_0 \varphi E^{(2)}(\partial u) + E^{(3)}(\partial u) + \mathcal{O}(\epsilon^4) \Big]\,,
\ee 
where $E^{(3)}(\partial u) = g_3 \text{Im}(\partial_z \partial_z \varphi)^3$ with $z=x+iy$ is a $C_6$-invariant cubic structure. More explicitly, the third-order action can be written in terms of the field $\varphi$ as $S[\varphi] = S_2[\varphi] + S_3[\varphi]$ with $S_2[\varphi]$ given by Eq. \eqref{eq:lifshitz} and
\be \label{eq:actionTkachenkoInteracting}
\begin{split}
S_3[\varphi] \simeq   \int d^3 x \Big[  g_1 (\partial_0 \varphi)^3 + \tilde g_2 \partial_0 \varphi  (\partial_a \partial_b \varphi)^2 - \bar g_2 \partial_0 \varphi (\partial^2 \varphi)^2  + g_3 \text{Im}(\partial_z \partial_z \varphi)^3 + \mathcal{O}(\epsilon^4) \Big]\,,
\end{split} \ee 
where $\tilde g_2 =  \Theta^2 \left( g_2 \lambda_2  + \frac{n_0}{4} \right)$ and $ \bar g_2 = \Theta^2 \left(\frac{g_2 \lambda_2}{2}  + \frac{n_0}{4} \right)$. 

The form of the action closely resembles the result of \cite{PhysRevResearch.6.L012040}. Upon comparison, however, our theory contains an extra term of the form 
\be \label{eq:additional} \partial_0 \varphi (\partial_a \partial_b \varphi)^2.
\ee
This term is allowed by symmetries, but was missed in \cite{PhysRevResearch.6.L012040}. Modulo this term, this agreement may seem surprising, since the authors of \cite{PhysRevResearch.6.L012040} work in the LLL limit, whereas our construction does not assume this limit. However, as emphasized around Eq. \eqref{eq:LLL}, beyond-LLL corrections are suppressed by higher-order contributions and do not affect this order in the expansion. Therefore, our results extend the validity of the LLL effective theory of \cite{PhysRevResearch.6.L012040}, (augmented by the missed interaction term), beyond the LLL approximation.

\section{Discussion and outlook}
In this work, we have developed an effective field theory for vortex crystals in rotating Bose-Einstein condensates by identifying  and systematically implementing the relevant spacetime and internal symmetries. We have verified that our construction satisfies several phenomenological constraints, including Kohn’s theorem and the presence of transverse Tkachenko oscillations. In writing down the most general nonlinear realization of the symmetry-breaking pattern, we encountered a number of redundant Goldstone fields, which can be eliminated using inverse Higgs constraints, to which we provided a clear physical interpretation. In particular, we argued that the Kohn mode is physical--despite being gapped--because it generates independent local fluctuations of the order parameter, in contrast to the boost and rotational Goldstones, which correspond to gauge redundancies. Finally, we eliminated the Kohn mode from the theory, imposed a consistent derivative expansion, and studied the resulting effective gapless theory.

In the concluding remarks, we discuss the comparison between our findings and previous works and outline several interesting directions for further investigation.

An effective Goldstone theory for the vortex lattice have been first derived directly from the underlying microscopic model \eqref{eq:rotatingFrame} by Watanabe and Murayama \cite{Watanabe:2013iia}. In contrast, our construction does not rely on the specific microscopic model but rather provides the most general effective theory consistent with the symmetries. While the leading-order theories coincide, this agreement does not extend beyond leading order.

The application of the coset construction to vortex crystals was initiated by Brauner and Watanabe, who also provided a physical interpretation of the inverse Higgs constraint associated with the Kohn mode \cite{PhysRevD.89.085004}. However, this construction was based only on the subset of the relevant spacetime symmetries and consequently it does not accurately generate the relevant low-energy building blocks. 

After eliminating the Kohn mode via its inverse Higgs constraint, we find that the resulting gapless theory is in agreement with the noncommutative field theory of \cite{PhysRevResearch.6.L012040} up to a third order in the expansion. However, as pointed out in the main text, we have identified an additional term Eq. \eqref{eq:additional} that is consistent with the symmetries but was missed in that work. Importantly, the authors of \cite{PhysRevResearch.6.L012040} work in the strict LLL limit ($m \rightarrow 0, \Omega \rightarrow \infty$), whereas our construction applies more generally for finite values of $\Omega$ and $m$.

Our results open up several promising directions for future research. First, it would be interesting to generalize our construction to different trapping potentials, including quartic traps and anisotropic potentials, which are relevant in many experimental configurations \cite{PhysRevLett.86.4443,PhysRevLett.94.150401,PhysRevA.79.011603,PhysRevLett.92.050403}. Incorporating finite temperature effects is another natural extension, enabling a systematic account of thermal fluctuations and dissipation. Finally, the inclusion of quantum corrections, through loop effects or renormalization group analysis, could shed light on the stability and universality of the low-energy effective theory, and on the role of quantum fluctuations in the dynamics of vortex crystals.

\begin{acknowledgments}
\noindent We thank P. Łydżba for useful discussions. A.G. acknowledges the visiting PhD fellow program at Nordita for funding a part of this work, and thanks Nordita for their hospitality. A.G. and P.S are supported in part by the Polish National Science Centre (NCN) Sonata Bis grant 2019/34/E/ST3/00405. S.M. is supported by Vetenskapsr{\aa}det (2021-03685), Carl Tryggers Stiftelse (CTS 24:3607), Wenner-Gren Stiftelserna (UPD2024-0111) and Nordita. F.P-B is supported by the NCN Sonata Bis grant 2024/54/E/ST2/00481.
\end{acknowledgments}

\appendix

 \section{Killing conditions}\label{app:rotatingKilling}
In this appendix, we present a detailed derivation of the Killing symmetry variations for Newton–Cartan geometry \eqref{eq:rotatingNC}, which we recall here for the reader’s convenience
\be
    \tau_{\mu} = \delta^0_\mu\,, \quad v^\mu = \delta^\mu_0\,, \quad h^{\mu \nu} = \delta^\mu_i \delta^\nu_j \delta^{ij}\,, \quad A_0 = \frac{1}{2} (\Omega^2-\omega^2) x^2\,, \quad A_i = \Omega \epsilon_{ij} x_j\,. 
\ee  
We have set $m=1$ for notational brevity but the factors of $m$ are easily restored by multiplying the $U(1)$ transformations by $m$. Using \eqref{eq:infitesimal} we find the following expressions for the symmetry transformations with parameters $\chi =(\xi^\mu \partial_\mu\,, \psi_\mu dx^\mu \,, \Lambda)$ of the geometric data
  \be \begin{split}
  \delta_\chi \tau_0 &=  \partial_0 \xi^0   \,, \\
    \delta_\chi \tau_i &=  \partial_i \xi^0   \,, \\
  \delta_\chi h^{0i} &=    -  \partial_i \xi^0  \,, \\
\delta_\chi h^{ij} &=   - \partial_j \xi^i -  \partial_i \xi^j  \,, \\
\delta_\chi v^0 &= -  \partial_0 \xi^0 \,, \\
\delta_\chi v^i &= -  \partial_0 \xi^i + \psi_i\,, \\
\delta_\chi A_0 & =  \xi^j \partial_j A_i + A_j \partial_i \xi^j + \dot \xi^i + \partial_i \Lambda \,, \\
\delta_\chi A_i & =  \xi^j \partial_j A_0 + A_i \dot \xi^i + \dot \Lambda  \,.
\end{split}\ee 
Imposing the conditions for vanishing $\delta_\chi \tau_0, \delta_\chi \tau_i, \delta_\chi h^{0i}$ and $\delta_\chi v^0$ imply that $\xi^0$ is a constant, which we denote $c^0 \in \mathbb R$. Then, the remaining independent Killing equations are 
  \be \begin{split}
\delta_\chi h^{ij} &=   - \partial_j \xi^i -  \partial_i \xi^j  =0\,, \\
\delta_\chi v^i &= -  \partial_0 \xi^i + \psi_i= 0\,, \\
\delta_\chi A_0 & =  \xi^j \partial_j A_i + A_j \partial_i \xi^j + \dot \xi^i + \partial_i \Lambda = 0\,, \\
\delta_\chi A_i & = \xi^j \partial_j A_0 + A_i \dot \xi^i + \dot \Lambda = 0\,.
\end{split}\ee 
The most general solution to the condition $\delta_\chi h^{ij} = 0$ is 
\be \label{eq:solution}
\xi^i = c^i(t) + \alpha(t) \epsilon_{ij} x_j \,.
\ee 
On the other hand, after eliminating $\psi_i=\partial_0 \xi^i$ from the condition for $\delta_\chi v^i=0$ we are left with two differential equations 
\be\begin{split}\label{eq:killings}
\xi^j \partial_j A_i + A_j \partial_i \xi^j + \dot \xi^i + \partial_i \Lambda = 0\,, \\
\xi^j \partial_j A_0 + A_i \dot \xi^i + \dot \Lambda = 0\,.
\end{split}
\ee 
Substituting the solution \eqref{eq:solution} into the first equation we find that 
\be
\Omega \epsilon_{ij} c_j(t)  + \partial_i \Lambda + \dot \alpha(t) \epsilon_{ij} x_j + \dot c^i(t) =0\,.
\ee 
Since the $x$--dependent term $\dot \alpha(t) \epsilon_{ij} x_j$ cannot be written as a total derivative we find that it must vanish and hence $\alpha(t)$ is time independent $\dot \alpha(t) = 0 \Longrightarrow \alpha \in \mathbb R$. Then, this equation also implies that 
\be 
\Lambda = \Omega \epsilon_{ij} x_j c^i(t) - x_i \dot c^i(t) + \lambda(t)\,,
\ee 
where $\lambda(t)$ is an arbitrary function of time only. Substituting this into the second equation in \eqref{eq:killings} we find
\be
\dot \lambda  =  x_i \Big( \ddot c^i(t) + 2\Omega \epsilon_{ij} \dot c_j(t)  + \big(\omega^2-\Omega^2\big) c^i(t) \Big) = 0\,,
\ee 
where in the second step we have used \eqref{eq:killingCoordinate}. Thus, we find that $\lambda(t) = \lambda$ is a time-independent constant.

It remains to find out the most general form of $c^i(t)$. Combining the two equations \eqref{eq:killings} we obtain a set of ODEs for $c^i(t)$ as follows
\be \label{eq:killingCoordinate}
\ddot c^i(t) + 2\Omega \epsilon_{ij} \dot c_j(t)  + \big(\omega^2-\Omega^2\big) c^i(t)  = 0 \,.
\ee
The most general solution is given in terms of four parameters 
\be
c^i(t) = -\frac{ \sin{(\omega t)}}{\omega} R_{ij}(\Omega t) b^0_j  - \cos{\big( \omega t\big)} R_{ij}(\Omega t) c^0_j- \frac{\Omega}{\omega} \sin{\big( \omega t \big)} R_{ik}\big(\Omega t \big) \epsilon_{kj} c^0_j\,,
\ee 
where $R_{ij}(\alpha) = \delta_{ij} \cos{(\alpha)}-\epsilon_{ij} \sin{(\alpha)}$ is an $SO(2)$ rotation matrix. 
Using $\psi^c_i(t) = \dot c_i(t)$ and $\Lambda^c(t) =\Omega \epsilon_{ij} x_j c_i(t) - x_i \dot c^i(t)$ we find the associated Milne boost and $U(1)$ components 
\be \begin{split}
\psi^c_i(t) &= -\cos{(\omega t)} R_{ij}(\Omega t) b^0_j +  \frac{\Omega }{\omega}  \sin{(\omega t)} R_{ik}(\Omega t) \epsilon_{kj} b^0_j+  \frac{ (\omega^2 - \Omega^2)}{\omega}  \sin{(\omega t)} R_{ij}(\Omega t) c_j^0 \,, \\
\Lambda^c(t) &= \cos{(\omega t)} R_{ij}(\Omega t) x_i b_j^0 +\Omega \cos{(\omega t)} R_{ik}(\Omega t) \epsilon_{kj} x_i c_j^0-\omega \sin{(\omega t)} R_{ij}(\Omega t)x_i c_j^0\,.
\end{split}
\ee

\section{Symmetry realizations}
In this appendix, we present realizations of the magnetic Bargmann \eqref{eq:magneticBargmann} and Newton-Hooke \eqref{eq:newtonHooke} symmetry groups in physical systems, both at the level of single-particle actions and in many-body (fluid) systems.

For single particle actions, we will employ the covariant fomulation of an action for the particle propagating on the Newton--Cartan spacetime \cite{Matus:2024kyg}
\be\label{eq:actionParticle}
L = \frac{m}{2 \tau_\rho \dot x^\rho} h_{\mu\nu} \dot{x}^\mu \dot{x}^\nu +  A^\mu \dot{x}_\mu\,.
\ee
On the other hand, the fluid realizations are based on the Euler equations in the presence of external electromagnetic forces.
 
\subsection{Magnetic Bargmann}\label{app:magneticBargmann}
\subsubsection{A particle in a magnetic field}
First, we consider a point particle moving in a homogeneous magnetic field and verify that the symmetry group of this system is the magnetic Bargmann group, whose generators satisfy the algebra \eqref{eq:magneticBargmann} with respect to the canonical Poisson bracket.

The single-particle Lagrangian is readily obtained by substituting the magnetic background \eqref{eq:rotatingNC} into the covariant action \eqref{eq:actionParticle}
\begin{equation}\label{eq:lagrangianParticleMagnetic}
L =  \frac{m}{2}  (\dot x_i)^2 + m\Omega \epsilon_{ij} \dot x_i x_j  \,.
\end{equation}
 Notice that this Lagrangian describes a particle moving in an effective magnetic field $B=-2m\Omega$. Indeed, in a symmetric gauge $A_i=-\frac{B}{2} \epsilon_{ij}x_j$, the Lagrangian is 
\be 
L =  \frac{1}{2} m (\dot x_i)^2 - \frac{1}{2}  B \epsilon_{ij}  \dot x_i x_j = \frac{m}{2}  (\dot x_i)^2 + m\Omega \epsilon_{ij} \dot x_i x_j \,.
\ee 
The equation of motion is the cyclotron equation
\begin{equation}\label{eq:eoms}
     \ddot x_i + 2\Omega \epsilon_{ij} \dot x_j =0\,.
\end{equation}
 In the context of a single particle whose trajectory is specified by $x_i(t)$, a symmetry is realized as an infinitesimal transformation of the coordinates $(t,x_i,\dot x_i) \rightarrow (t +\delta t, x_i + \delta x^i, \dot x_i +\delta \dot x_i)$ provided that it leaves the Lagrangian invariant up to a total time derivative $\delta L = \frac{dF}{dt}$ where $F\equiv F(t,x_i, \dot x_i)$ is an arbitrary function. 

It is straightforward to verify that the theory \eqref{eq:lagrangianParticleMagnetic} enjoys the following symmetries:
\be
\begin{aligned}\label{eq:particleSymmetries}
\delta x_i &= c_0 \dot x_i\,, 
&\quad \delta \dot x_i &= c_0 \ddot x_i\,,
&\quad F &= c_0 \mathcal L\,, \\
\delta x_i &= -\alpha \epsilon_{ij} x_j\,,
&\quad \delta \dot x_i &= -\alpha \epsilon_{ij} \dot x_j\,,
&\quad F &= 0\,, \\
\delta x_i &= -\frac{1}{2\Omega} \epsilon_{ij} \left( b_j^0 - R_{jk}(2\Omega t) b_k^0 \right)\,,
&\quad \delta \dot x_i &= R_{ij}(2\Omega t) b_j^0\,,
&\quad F &= \frac{m}{2}  x_i R_{ij}(2\Omega t) b_j^0 + \frac{m}{2} x_i b_i^0, \\
\delta x_i &= c^0_i\,,
&\quad \delta \dot x_i &= 0\,,
&\quad F &= - m \Omega \epsilon_{ij} x_j c^0_i\,.
\end{aligned}
\ee
By the Noether's theorem, for each symmetry \eqref{eq:particleSymmetries} there exists a conserved charge 
\begin{equation}
   Q = \frac{\partial L}{\partial \dot x_i} \delta x_i - F\,,
\end{equation}
where the conjugate momentum is $\pi_i = \frac{\partial L}{\partial \dot x_i} = m \dot x_i+m\Omega \epsilon_{ij}x_j$. Therefore, we identify the following conserved quantities 
\begin{equation}
    \begin{split}
    H & = \frac{1}{2m} \big( \pi_i - m \Omega \epsilon_{ij} x_j \big)^2 \,, \\
        J &= -\epsilon_{ij}  \pi_i x_j  \,, \\
            P_i & = \pi_i + m\Omega \epsilon_{ij}x_j \,, \\
    B_i &= \frac{1}{2\Omega} \epsilon_{ij} \big(\pi_j+m\Omega\epsilon_{jk}x_k \big) - \frac{1}{2\Omega} R_{ij}(-2\Omega t) \epsilon_{jk} \big(\pi_k - m\Omega \epsilon_{kl} x_l  \big) \,.
    \end{split}
\end{equation}
Imposing the canonical Poisson bracket $\{x_i, \pi_j \} = \delta_{ij}$ we find that the conserved quantities satisfy the magnetic Galilean algebra
\be \label{eq:magneticGalileanAlgebra}
\{P_i, P_j \} = -2m\Omega \epsilon_{ij} \,, \quad \{H, B_i\} = -P_i - 2\Omega \epsilon_{ij} B_j\,, \quad \{B_i, P_j\} = m \delta_{ij}\,, \quad \{J, P_i\} = \epsilon_{ij} P_j\,, \quad \{J, B_i\} = \epsilon_{ij} B_j \,.
\ee

\subsubsection{Lowest Landau Level limit}\label{sec:LLL}

We now illustrate the claim regarding the Lowest Landau Level (LLL) limit. To this end, it is useful to perform a change of variables from the configuration space coordinates \((x_i, \dot{x}_i)\) to the guiding center and cyclotron coordinates \((X_i, V_i)\), defined as
\begin{equation}
    X_i = \pi_i + m \Omega \epsilon_{ij} x_j \,, \quad V_i = \pi_i - m \Omega \epsilon_{ij} x_j \,.
\end{equation}
These coordinates satisfy the commutation relations
\begin{equation}
    [X_i, V_j] = 0 \,, \quad [X_i, X_j] = 2 m \Omega \epsilon_{ij} \,, \quad [V_i, V_j] = -2 m \Omega \epsilon_{ij} \,.
\end{equation}

It is then convenient to perform a change of Lie algebra basis by introducing the generator
\begin{equation}
    K_i = B_i + \frac{1}{2\Omega} \epsilon_{ij} P_j \,.
\end{equation}
In terms of \(K_i\), the algebra \eqref{eq:magneticGalileanAlgebra} takes the form
\begin{equation}
    \{P_i, P_j\} = -2 m \Omega \epsilon_{ij} \,, \quad
    \{K_i, K_j\} = -2 \frac{m}{\Omega} \epsilon_{ij} \,, \quad
    \{H, K_i\} = -2 \Omega \epsilon_{ij} K_j \,, \quad
    \{J, P_i\} = \epsilon_{ij} P_j \,, \quad
    \{J, K_i\} = \epsilon_{ij} K_j \,.
\end{equation}

Notice that the infinitesimal action of \(K_i\) on the coordinates \((x_i, \dot{x}_i)\) is given by
\begin{equation}
    e^{K_i d^0_i} (x_i, \dot{x}_i) \approx \left( x_i + \frac{1}{2\Omega} \epsilon_{ij} R_{jk}(2\Omega t) d^0_k, \quad \dot{x}_i + R_{ij}(2\Omega t) d^0_j \right) \,.
\end{equation}
Consequently, the generators \(P_i\) and \(K_i\) act on the guiding center and cyclotron coordinates \((X_i, V_i)\) as
\begin{equation}
    e^{P_i c^0_i} e^{K_i d^0_i} (X_i, V_i) \approx \left( X_i + 2 m \Omega \epsilon_{ij} c^0_j, \quad V_i + m R_{ij}(2\Omega t) d^0_j \right) \,.
\end{equation}
Therefore, \(P_i\) and \(K_i\) act nontrivially only on the guiding center and cyclotron coordinates, respectively.

In the Lowest Landau Level, the cyclotron coordinate is frozen \cite{doi:10.1126/science.aba7202}, i.e., \(V_i = m \dot{x}_i \rightarrow 0\), and the dynamics is fully captured by the guiding center coordinate \(X_i\). Since the Hamiltonian $H = \frac{1}{2m} (V_i)^2 \rightarrow 0\,,$ it is also frozen.

Thus, in the Lowest Landau Level the only dynamical degrees of freedom are the guiding center coordinates \(X_i\), and the relevant symmetry group acting nontrivially is fully captured by the magnetic translation algebra
\begin{equation}
    \{P_i, P_j\} = -2 m \Omega \epsilon_{ij} \,, \quad \{J, P_i\} = \epsilon_{ij} P_j \,.
\end{equation}

\subsubsection{Charged fluid dynamics in a magnetic field}
 In this section of the appendix we consider a realization of the magnetic Bargmann group in charged fluids placed in a homogeneous magnetic field (we set $B=-2m\Omega$).

 The dynamics of such a system is governed by the hydrodynamic equations
 \be \begin{split}
 \partial_t \rho + \partial_i (\rho v_i) &= 0\,, \\
  \partial_t v_i + v_j \partial_j v_i  &= -\frac{1}{\rho} \partial_i p - 2\Omega \epsilon_{ij} v_j\,,
  \end{split}\ee 
where $\rho = m n$ is mass density and the last term represents the contribution from the Lorentz (or Coriolis) force. 

These equations are captured by the Hamiltonian
\be \label{eq:magneticFluid}
H = \int d^2 x \Big[ \frac{\rho}{2} v^2  + V(\rho) \Big]\,,
\ee 
supplemented with the Poisson bracket structure
\begin{align}
\{  v_i(\textbf x), \rho(\textbf y) \} &= -\partial_i \delta(\textbf x - \textbf y) \,, \\
\{ v_i(\textbf x), v_j(\textbf y) \} &=  \frac{1}{\rho(\mathbf x)} \big[ \omega_{ij} - 2\Omega \epsilon_{ij}\big] \delta(\textbf x - \textbf y) \,,
\end{align}
where $\omega_{ij}=\partial_i v_j(\mathbf x)-\partial_j v_i(\mathbf x)$ is fluid's vorticity. 

This can be verified via a direct computation
\be\begin{split}
\partial_t \rho &= \{\rho, H\} = -\partial_i (\rho v_i )\,, \\
\partial_t v_i &= \{v_i, H\} =  -\partial_i V^\prime(\rho) - v_j \partial_j v_i - 2\Omega \epsilon_{ij} v_j \,. 
\end{split}
\ee 

Having recast the dynamics in the Hamiltonian framework, it is straightforward to verify that the system admits the following set of conserved quantities
\be \begin{split}
N &= \int d^2 x \, n \,, \\
H &= \int d^2 x \Big[ \frac{\rho}{2} v^2  + V(\rho) \Big]\,, \\
L &= \int d^2 x \, \Big[ \epsilon_{ij} \rho v_j x_i - \rho \Omega x^2 \Big] \,, \\
P_i &= \int d^2 x \, \Big[ \rho v_i  +2\rho \Omega \epsilon_{ij} x_j\Big]\,, \\
B_i &= \int d^2 x \, \frac{1}{2\Omega}\Big[\epsilon_{ij}  \big( \rho v_j+2\rho\Omega \epsilon_{jk} x_k\big) - R_{ij}(-2\Omega t) \epsilon_{jk} \rho v_k \Big] \,.
\end{split}
\ee 
 It is also possible (although quite tedious) to check that the conserved charges satisfy the magnetic Bargmann algebra 
\be 
\{P_i, P_j \} = -2\Omega \epsilon_{ij} m N \,, \quad \{H, B_i\} = -P_i - 2\Omega \epsilon_{ij} B_j\,, \quad \{B_i, P_j\} = \delta_{ij} m N\,, \quad \{L, P_i\} = \epsilon_{ij} P_j\,, \quad \{L, B_i\} = \epsilon_{ij} B_j \,.
\ee 

\subsection{Newton-Hooke}\label{app:NewtonHooke}
In this section, we discuss the Newton–Hooke symmetry group \eqref{eq:newtonHooke} in the classical harmonic oscillator and ideal fluid dynamics within a harmonic trap.

\subsubsection{Harmonic oscillator}
To obtain a single particle realization of the Newton-Hooke symmetry, we plug the background \eqref{eq:harmonicNC} into the covariant Lagrangian \eqref{eq:actionParticle}. After doing so, we arrive at the well-known harmonic oscillator theory 
\be 
L = \frac{1}{2} m \dot x^2-\frac{1}{2} m\omega^2 x^2\,.
\ee
It is straightforward to check that the harmonic oscillator theory possesses the following symmetries
\be
\begin{aligned}\label{eq:particleSymmetriesNewtonHooke}
\delta x_i &= c_0 \dot x_i\,, 
&\quad \delta \dot x_i &= c_0 \ddot x_i\,,
&\quad F &= c_0 \mathcal L\,, \\
\delta x_i &= -\alpha \epsilon_{ij} x_j\,,
&\quad \delta \dot x_i &= -\alpha \epsilon_{ij} \dot x_j\,,
&\quad F &= 0\,, \\
\delta x_i &= b^0_i \frac{\sin{(\omega t)}}{\omega} \,,
&\quad \delta \dot x_i &= b^0_i  m   x_i \cos{(\omega t)} \,,
&\quad F &= b^0_i  m x_i \cos{(\omega t)}, \\
\delta x_i &= c_i^0 \cos{(\omega t)} \,,
&\quad \delta \dot x_i &= -c_i^0 \omega \sin{(\omega t)}\,,
&\quad F &= - c^0_i  m \omega x_i \sin{(\omega t)}\,.
\end{aligned}
\ee
Applying the Noether's theorem, we identify the conserved quantities  
\begin{equation}
    \begin{split}
    H & = \frac{1}{2} m \dot x^2 + \frac{1}{2} m\omega^2 x^2\,, \\
    B_i &= m\dot x_i \frac{\sin{(\omega t)}}{\omega} - m x_i \cos{(\omega t)}\,, \\
        P_i & = m\dot x_i \cos{(\omega t)} + m \omega x_i \sin{(\omega t)}\,, \\
    J &= m \epsilon_{ij}  \dot x_i  x_j \,.
    \end{split}
\end{equation}

After introducing the canonical Poisson bracket we find that the conserved quantities satisfy the  following algebra 
\be 
\{B_i, P_j \} = m \delta_{ij}\,, \quad \{H, B_i\} = P_i\,, \quad \{H, P_i\} = -\omega^2 B_i\,, \quad \{L, P_i\} = \epsilon_{ij} P_j\,, \quad \{L, B_i\} = \epsilon_{ij} B_j \,.
\ee 

\subsubsection{Charged fluid dynamics on a harmonic trap}
Another realization of the Newton-Hooke symmetry group is given by ideal charged fluid dynamics in a harmonic trap \cite{SNEGIREV2025116902}. We start with the Hamiltonian  
\be 
H = \int d^2 x \Big[ \frac{1}{2}\rho v^2  + \frac{1}{2} \rho \omega^2 x^2 + V(\rho) \Big]\,,
\ee 
where $V\equiv V(\rho)$ is the intermolecular potential. In this case, the hydrodynamic variables satisfy the canonical Poisson brackets
\begin{align}
\{  v_i(\textbf x), \rho(\textbf y) \} &= -\partial_i \delta(\textbf x - \textbf y) \,, \\
\{ v_i(\textbf x), v_j(\textbf y) \} &=  \frac{1}{\rho(\mathbf x)} \omega_{ij}  \delta(\textbf x - \textbf y) \,. 
\end{align}
The Euler equations are recovered via
\be\begin{split}
\partial_t \rho &= \{\rho, H\} = -\partial_i (\rho v_i )\,, \\
\partial_t v_i &= \{v_i, H\} =  -\partial_i V^\prime(\rho) - v_j \partial_j v_i - \omega^2 x_i \,. 
\end{split}
\ee 

It is then possible to construct the following conserved quantities 
\be \begin{split}
N &= \int d^d x \, \rho \,, \\
H &= \int d^d x \, \Big[ \frac{1}{2}\rho v^2  + \frac{1}{2} \rho \omega^2 x^2 + V(\rho) \Big] \,, \\
L &= \int d^d x \, \Big[\epsilon_{ij} \rho v_i x_j \Big] \,, \\
P_i &= \int d^d x \, \Big[ \rho v_i \cos{(\omega t)} + m \omega x_i \sin{(\omega t)}\Big] \,, \\
B_i &= \int d^d x \, \Big[\rho v_i \frac{\sin{(\omega t)}}{\omega} - m x_i \cos{(\omega t)} \Big] \,.
\end{split}
\ee 
These conserved quantities satisfy the Newton-Hooke algebra with respect to the canonical Poisson bracket structure 
\be 
\{B_i, P_j \} = \delta_{ij} mN \,, \quad \{H, B_i\} = P_i\,, \quad \{H, P_i\} = -\omega^2 B_i\,, \quad \{L, P_i\} = \epsilon_{ij} P_j\,, \quad \{L, B_i\} = \epsilon_{ij} B_j \,.
\ee

\bibliography{biblio}

\end{document}